\documentclass[pdflatex,sn-nature,iicol]{sn-jnl}
\usepackage{graphicx}%
\usepackage{multirow}%
\usepackage{amsmath,amssymb,amsfonts}%
\usepackage{amsthm}%
\usepackage{mathrsfs}%
\usepackage{xcolor}%
\usepackage{soul}
\usepackage{gensymb}
\usepackage{dcolumn}
\usepackage{xfrac}
\usepackage{bm}
\newcolumntype{d}[1]{D{.}{.}{#1}}

\definecolor{go_green}{rgb}{0.13, 0.55, 0.13}

\usepackage[normalem]{ulem}

\makeatletter
\begin{document}
	\title[]{Symmetry tuning topological states of the axion insulator candidate EuIn$_2$As$_2$ by its noncollinear magnetic order}
	
	\author[1,2]{\fnm{S.~X.~M.}~\sur{Riberolles}}
	
	\author[1,2]{\fnm{A.-M.}~\sur{Nedi\'c}}
	
	\author[1,2]{\fnm{B.}~\sur{Kuthanazhi}}
	
	\author[3]{\fnm{F.}~\sur{Ye}}
	
	\author[1,2]{\fnm{S.~L.}~\sur{Bud'ko}}
	
	\author[1,2]{\fnm{P.~C.}~\sur{Canfield}}
	
	\author[1,2]{\fnm{R.~J.}~\sur{McQueeney}}
	\author[4]{\fnm{Junyeong}~\sur{Ahn}}
	
	\author[1,2,5]{\fnm{V.~L.}~\sur{Quito}}
	
	\author[1,2,5]{\fnm{T.~V.}~\sur{Trevisan}}
	
	\author[1,2]{\fnm{L.~L.}~\sur{Wang}}
	
	\author*[1,2,6]{\fnm{P.~P.}~\sur{Orth}}
	\email{peter.orth@uni-saarland.de}
	
	\author*[1,2]{\fnm{B.~G.}~\sur{Ueland}}
	\email{bgueland@ameslab.gov}
	
	\affil[1]{\orgdiv{Division of Materials Sciences and Engineering}, \orgname{Ames National Laboratory, U.S. DOE, Iowa State University}, \orgaddress{ \city{Ames}, \state{IA}, \country{USA}, \postcode{50011}}}
	
	\affil[2]{\orgdiv{Department of Physics and Astronomy}, \orgname{Iowa State University}, \orgaddress{ \city{Ames},  \state{IA}, \country{USA}, \postcode{50011}}}
	
	\affil[3]{\orgdiv{Neutron Scattering Division}, \orgname{Oak Ridge National Laboratory}, \orgaddress{ \city{Oak Ridge}, \state{TN},  \country{USA}, \postcode{37831}}}
	
	\affil[4]{\orgdiv{Department of Physics}, \orgname{Harvard University}, \orgaddress{ \city{Cambridge}, \state{MA},  \country{USA}, \postcode{02138}}}
	
	\affil[5]{\orgdiv{Sao Carlos Institute of Physics}, \orgname{University of Sao Paulo, IFSC – USP},\orgaddress{  \city{Sao Carlos}, \state{SP}, \country{BR}, \postcode{13566-590}}}
	
	\affil[6]{\orgdiv{Department of Physics}, \orgname{Saarland University},\orgaddress{  \city{Saarbr\"{u}cken}, \country{DE}, \postcode{66123}}}
	
	\date{\today}	
	
	\abstract{Topological properties of quantum materials are intimately related to symmetry. Here, we tune the magnetic order of the axion insulator candidate EuIn$_2$As$_2$ from its broken-helix ground state to the field-polarized phase by applying an in-plane magnetic field. Using results from neutron diffraction and magnetization measurements with ab inito theory and symmetry analysis, we assume a commensurate magnetic ground state and predict how the field tunes the magnetic symmetry within individual magnetic domains and examine the resulting changes to the topological surface states and hinge states existing on edges shared by certain surfaces hosting gapped Dirac states. We predict complex field-tunable domain-specific hinge-state patterns, with some crystal surfaces undergoing a field induced topological phase transition. We further find that domain walls can have pinned hinge states when intersecting certain crystal surfaces, providing another channel for tuning the chiral-charge-transport pathways.}
	
	
	\maketitle
	
	\footnotetext[1]{A.-M.~Nedi\'{c} Present Address: Chemical Engineering and Materials Science, University of Minnesota, Minneapolis, MN 55455, USA}
	
	\footnotetext[2]{B.~Kuthanazhi Present Address: Department of Chemistry, University of Liverpool, Liverpool, L69 7ZD, UK}
	
	\footnotetext[3]{Junyeong~Ahn Present Address: Department of Physics, The University of Texas at Austin, Austin, TX 78712}
	
	\section{Introduction}
	
	Quantum materials with nontrivial topology can have topologically protected boundary states offering useful functional properties such as dissipationless and spin-polarized transport, quantized conductivity, or novel magnetoelectricity \cite{Hasan_2010, Moore_2010,Hasan_2011, Vanderbilt_2018, Chang_2023}. The existence and classification of a material's topology is intimately related to its symmetry. Therefore, the application and tuning of symmetry-breaking perturbations offer different pathways for controlling topological states and their robust physical properties. Here, we examine how the magnetic symmetry of the axion insulator (AXI) candidate EuIn$_2$As$_2$ could be tuned by an in-plane magnetic field from a complex helical ground state to its field-polarized phase and analyze how the changing symmetry controls the topologically protected boundary states.
	
	\begin{figure*}
		\centering
		\includegraphics[width=1\linewidth]{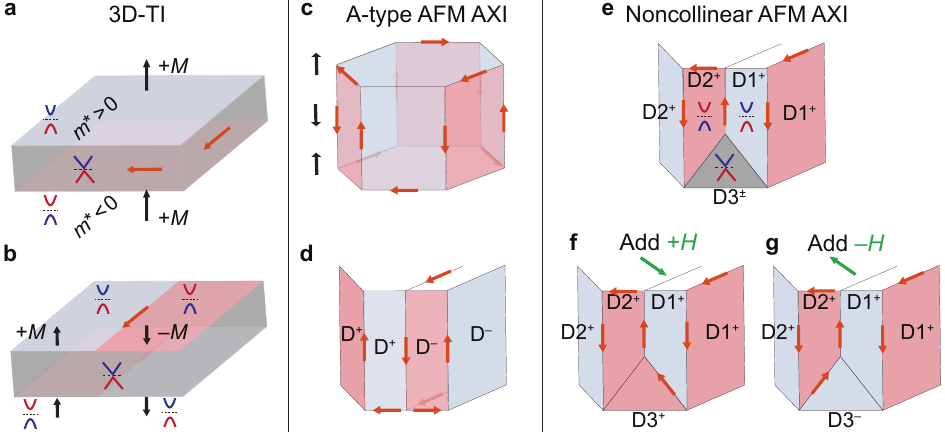}%
		\caption{
			\textbf{Examples of magnetically tuning bulk topological insulators.} \textbf{a} A bulk ($3$D) topological insulator (TI) between two ferromagnets where a magnetization $M$ gaps the Dirac states of the top and bottom surfaces. Red and blue surfaces have gapped Dirac states with opposite signs for the effective mass $m^*$. Light-gray surfaces have gapless Dirac states, and orange arrows indicate chiral-conduction paths. \textbf{b} $M$ changes sign across the top and bottom surfaces at a magnetic domain wall. A chiral-conduction path is pinned to the domain wall. \textbf{c} An axion insulator (AXI) with A-type antiferromagnet (AFM) order as indicated by the black arrows. All surfaces are gapped, and inversion-related surfaces have opposite signs for $m^*$. Hinge states run between red and blue surfaces. \textbf{d} Two domains (D$^+$ and D$^-$) related by $\mathcal{T}$ intersect at a magnetic domain wall on the surface of the A-type ordered AXI. A hinge state occurs where the domain wall intersects the surface. \textbf{e} Three magnetic domains intersecting on a surface of an AXI with complex (e.g.\ noncollinear) magnetic order. Domains D$1^+$ and D$2^+$ have gapped surface Dirac states with opposite $m^*$, but domain D$3^\pm$ has gapless surface Dirac states. \textbf{f}, \textbf{g} A magnetic field $H$ can gap the surface Dirac states of D$3^{\pm}$ and the direction of $\mathbf{H}$ chooses the chirality ($+$ or $-$) and the sign of $m^*$. Thus, the field direction selects whether a hinge state exists on the wall between D$1^+$ and D$3^+$ (f) or between D$2^+$ and D$3^-$ (g).   
		}
		\label{Fig:Examples}
	\end{figure*}
	
	To begin, we present some examples demonstrating the relationship between symmetry and topological boundary states, which will be important for understanding our results for EuIn$_2$As$_2$. Figure~\ref{Fig:Examples}a shows a bulk (i.e.\ $3$D) topological insulator (TI) located between two ferromagnets with magnetization $M$ pointing up. In the absence of $M$, bulk-boundary correspondence leads to robust metallic surfaces, with each surface hosting an odd number of Dirac cones~\cite{Hasan_2010}. The presence of $M$, however, acts as a local time-reversal-symmetry ($\mathcal{T}$) breaking perturbation at the top and bottom surfaces. This gaps their Dirac states and their Dirac fermions acquire an effective mass $m^*$. $m^*$ has opposite sign for the top and bottom surfaces (represented by blue and red) because $\text{sgn}(m^*)\propto \mathbf{\hat{M}} \cdot \mathbf{\hat{n}}$, where $\mathbf{\hat{n}}$ is the surface normal.  The gapped top and bottom surfaces have topologically protected quantized-conductance-channels which intersect the metallic side surfaces.
	
	Topologically protected chiral-conduction channels are not constrained, however, to develop at crystalline terminations. Figure~\ref{Fig:Examples}b shows the case of ferromagnets with opposite signs for $M$ placed across both the top and bottom surfaces of the $3$D TI. For both surfaces, this generates regions with opposite signs for $m^*$ which are separated by a domain wall. A chiral-conduction pathway emerges at the domain wall. In general, engineering such local $\mathcal{T}$-breaking perturbations can be explored to create robust topologically protected conduction along desired pathways \cite{Qi_2008}. In particular, the exploitation of magnetic domains and domain walls has important applications in information processing and storage~\cite{Cowburn_Science_2008,Thomas_Science_2008,Vedmedenko2020}, and in spintronics~\cite{DasSarma_RMP_2004,Hillebrands_JMMM_2020}. The new degrees of freedom offered by topologically protected chiral-conduction states makes studying how to control magnetic symmetry and domains of a TI promising paths for discovery.
	
	Whereas the above examples relied on the proximity of a TI to ferromagnets, one can consider an intrinsically magnetic TI where the development of magnetic order lowers the symmetry and profoundly impacts the topological properties. Figure~\ref{Fig:Examples}c shows a hexagonal magnet which is a strong $3$D TI in its paramagnetic phase. Upon cooling below a N\'eel temperature $T_{\text{N}}$, it develops the A-type antiferromagnetic (AFM) order indicated by the black arrows representing planes of ferromagnetically aligned ordered magnetic moments (spins) stacked antiferromagnetically along the crystalline $\mathbf{c}$ axis. The spins are oriented perpendicular to the $\mathbf{ab}$ crystalline plane. The A-type order breaks $\mathcal{T}$ but preserves inversion symmetry ($\mathcal{P}$), which leads to the material being an AXI for the magnetically ordered phase. AXI is a remarkable topological phase of matter characterized by quantized bulk magnetoelectric coupling and half-quantized quantum-anomalous-Hall type conductivity for gapped surfaces~\cite{Vanderbilt_2018,wilczek1987two,Qi_2008}.
	
	Generally, an AXI is protected by any symmetry other than $\mathcal{T}$ alone that reverses the sign of a pseudoscalar, which in the case of the A-type AFM TI in Fig.~\ref{Fig:Examples}c is $\mathcal{P}$. However, $\mathcal{P}$ is broken at all crystallographic surfaces, and all of the surfaces shown are thus generically gapped. Importantly, $m^*$ has opposite signs for $\mathcal{P}$-related surfaces~\cite{Tanaka_2020}, and topologically protected hinge states develop at the edges shared by surfaces with opposite signs for $m^*$. This is shown by the orange arrows for the example pattern of $\mathcal{P}$-related surfaces given in Fig.~\ref{Fig:Examples}c. The hinge-state chiral-conduction pathways circulate clockwise around surfaces with $m^*>0$ and counterclockwise for $m^*<0$~\cite{Tanaka_2020}.
	
	Symmetry lowering by magnetic ordering can lead to the appearance of magnetic domains, making it important to understand the impact of domain formation on topological properties. As demonstrated in Fig.~\ref{Fig:Examples}b, topologically protected chiral states can be pinned to domain walls \cite{Wakatsuki_2015}. We also can consider the case of two AFM domains related by $\mathcal{T}$ in the A-type AFM AXI of Fig.~\ref{Fig:Examples}c. This is shown in Fig.~\ref{Fig:Examples}d where domains D$^+$ and D$^-$  are separated by a domain wall intersecting a front surface. Since $m^*$ changes sign under $\mathcal{T}$, surfaces connected by the domain wall will necessarily have opposite signs for $m^*$. This guarantees a hinge state developing at the intersection of the domain wall with a surface.
	
	More complex noncollinear magnetic ordering can host multiple domains related by a combination of spatial symmetry operations and $\mathcal{T}$. Figure~\ref{Fig:Examples}e shows an example where three domains intersect on a high-symmetry surface of an AXI with noncollinear magnetic order. The intersecting surfaces shown for domains D$1^+$ and D$2^+$ are gapped and have opposite signs for $m^*$, whereas D$3^\pm$ have gapless (metallic) surfaces. Thus, no hinge states are found at walls shared with D$3^\pm$. However,  when applying a magnetic field $H$, the field can affect the magnetic ordering of each domain differently. Therefore, the field also can impact the topological surface states of each domain differently. For instance, as shown in Figs.~\ref{Fig:Examples}e--\ref{Fig:Examples}g, a weak in-plane field can gap the surface Dirac states of D$3^\pm$ while keeping the signs of $m^*$ for D$1^+$ and D$2^+$ unchanged, with the direction of $\mathbf{H}$ controlling the appearance of new hinge states. It is worth emphasizing that such a scenario truly requires noncollinear magnetic order and is thus beyond the collinear case shown in Figs.~\ref{Fig:Examples}c and \ref{Fig:Examples}d. Most importantly, it allows tuning the location of hinge modes with small magnetic fields and over large distances \emph{without} moving magnetic domain walls, which is necessary in the collinear case~\cite{Yasuda_2017}. With this clear motivation in mind, we now turn to  EuIn$_2$As$_2$. 
	
	\begin{figure}	
		\centering
		\includegraphics[width=1\linewidth]{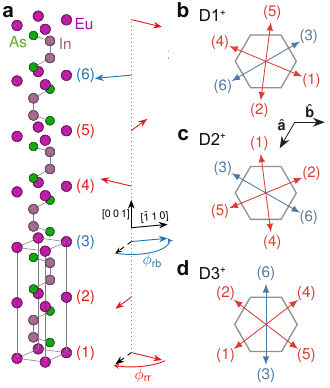}
		\caption{
			\textbf{Broken-helix magnetic order and magnetic domains for EuIn$_2$As$_2$.} \textbf{a} The chemical structure (left) of EuIn$_2$As$_2$ and broken-helix magnetic order (right) for domain D$1^+$  with helix angles  $\phi_{\text{rr}}$ and $\phi_{\text{rb}}$. The hexagonal chemical unit cell [$a=4.178(3)$~\AA\ and $c=17.75(2)$~\AA] is outlined. Red and blue arrows correspond to the ferromagnetically aligned magnetic moments (spins) lying in the $ab$ plane for each numbered Eu layer.  \textbf{b}--\textbf{d} Top-down views of the orientation of the layers in the $ab$ plane for domains D$1^+$ (b), D$2^+$ (c), and D$3^+$ (d), where $+$ denotes positive chirality. The domains are related by $120\degree$ rotations, and similar diagrams for the negative chirality domains would show an overall clockwise rotation of layers. The blue spins [layers ($3$) and ($6$)] lay along $\pm[\bar{1},1,0]$ for D$1^+$ (and D$1^-$).  
		}
		\label{Fig:Structure}
	\end{figure}
	
	Hexagonal EuIn$_2$As$_2$ (space group $P6_3/mmc$) has triangular magnetic Eu$^{2+}$ ($S=7/2$) layers intercalated with In$_2$As$_2$ blocks, as shown in Fig.~\ref{Fig:Structure}a. The In$_2$As$_2$ blocks have inverted electronic bands, leading to a nontrivial bulk topology. The Eu spins order below $T_{\text{N}}=16.2(1)$~K into the broken-helix magnetic structure shown in Fig.~\ref{Fig:Structure}a \cite{Riberolles_2021, Donoway_2023}; spins in each Eu layer are ferromagnetically aligned and oriented perpendicular to $\mathbf{c}$, with the direction of the spins changing between layers as illustrated. Here we assume that the magnetic unit cell is tripled along $c$, with two symmetry-inequivalent Eu sites colored either red [layers ($1$), ($2$), ($4$), and ($5$)] or blue [($3$) and ($6$)]. This structure is described by magnetic space group (MSG) $20.33$ which has $C2^\prime2^\prime2_1$ symmetry. The helix angles are $\phi_{rr}$ and $\phi_{rb}$, with the constraint $\phi_{rr}+2\phi_{rb}=180\degree$. We note that MSG $20.33$ implies the existence of six magnetic domains, generated by threefold rotation and inversion symmetry. In the discussion of our results below, we find that the existence of domains is supported by the observation of hysteresis and the fact that different magnetic Bragg peaks exhibit features at distinct magnetic field strengths, reflecting the domains' different in-field evolution.
	
	There is some controversy surrounding the $C2^\prime2^\prime2_1$ description of the magnetic order, as the ground-state AFM propagation vector is slightly incommensurate [$\bm{\tau_1}=(0,0,\tau_{1z})$, with $\tau_{1z} \approx 1/3$] \cite{Riberolles_2021, Soh_2023, Donoway_2023, Takeda_2024, Gen_2025, Singh_2025}. As we describe below, however, the incommensurability for our samples is quite small [$\tau_{1z} = 0.323(8)$] and can be treated as a small perturbation, corresponding to a disagreement in the magnetic state at every $100$th Eu layer, which is not expected to destroy the AXI phase. Further, the atypical nature of the broken-helix ordering for EuIn$_2$As$_2$ has been verified by others, where recent work has found results consistent with broken-helix ordering below $T_{\text{N}}$; an x-ray resonance magnetic scattering study found $\tau_{1z}=1/3$ within experimental error \cite{Soh_2023}; a multimodal study concluded that slight uniaxial strain is necessary to pin the blue spins to specific crystalline directions and protect an AXI phase \cite{Donoway_2023}. Finally, Eu$^{2+}$ generally behaves as a local moment within a compound and is therefore not expected to generate spin-density-wave type order at low temperature $T < T_{\text{N}}$. Our density-functional-theory (DFT) results as well as those reported by others indicate that the Eu $4f$ levels lay below the Fermi energy $E_{\text{F}}$ and the ordered moment is consistent with the value of $7~\mu_{\text{B}}$ expected for Eu$^{2+}$ \cite{Riberolles_2021, Sato_2020}. There have been proposals that mediation of the magnetism by holes may lead to an incommensurate value for $\tau_{1z}$ \cite{Gen_2025}.
	
	EuIn$_2$As$_2$ has been experimentally shown to be a slightly hole-doped compensated semi-metal. However, due to the robust quantization of the axion angle $\theta$ at $\theta=\pi$, the slight doping necessary to make it a bulk insulator is not expected to change the magnetic ground state significantly enough to destroy the AXI state \cite{Riberolles_2021}. Substituting Eu by Ca achieves electron doping and brings the Fermi surface closer to the topological band gap without altering the magnetic state \cite{Yan_2024}.
	
	The broken-helix order shown in Fig.~\ref{Fig:Structure} supports an AXI state protected by $2^\prime$ symmetry, which is the combination of a two-fold rotation and $\mathcal{T}$ \cite{Riberolles_2021, Donoway_2023}. Since the MSG is orthorhombic and the chemical lattice is hexagonal, six magnetic domains form. Figures~\ref{Fig:Structure}b--\ref{Fig:Structure}d show three of them (D$1^+$, D$2^+$, and D$3^+$), which are related by $120\degree$ rotations around $\mathbf{c}$. The superscript $+\text{/}-$ denotes chirality, and D$1^{-}$, D$2^{-}$ and D$3^{-}$ are related to their $+$ counterpart by $\mathcal{P}$. D$1^{\pm}$, D$2^{\pm}$, and D$3^{\pm}$ have distinct sets of $2^\prime$ symmetry axes related by $120\degree$ rotations, which imposes constraints on $m^*$ specific to surfaces for each domain. This is because surfaces related by $2^\prime$ are gapped and have opposite signs for $m^*$, whereas, surfaces normal to a $2^\prime$ axis are gapless.
	
	We next show magnetization and neutron diffraction data for a single crystal of EuIn$_2$As$_2$ under the influence of an in-plane magnetic field applied along $\mathbf{b}$. We semi-quantitatively analyze the data by deriving a symmetry-constrained model that gives the commensurate broken-helix ground state and calculate field-induced changes to the magnetic symmetry of each domain. We finish by using symmetry analysis to investigate some possible accompanying changes to the topological surface states and illustrate some of the complex field-induced hinge-state patterns.
	
	\section{Results}
	\subsection{Magnetization and Neutron Diffraction}
	\begin{figure*}
		\centering
		\includegraphics[width=1\linewidth]{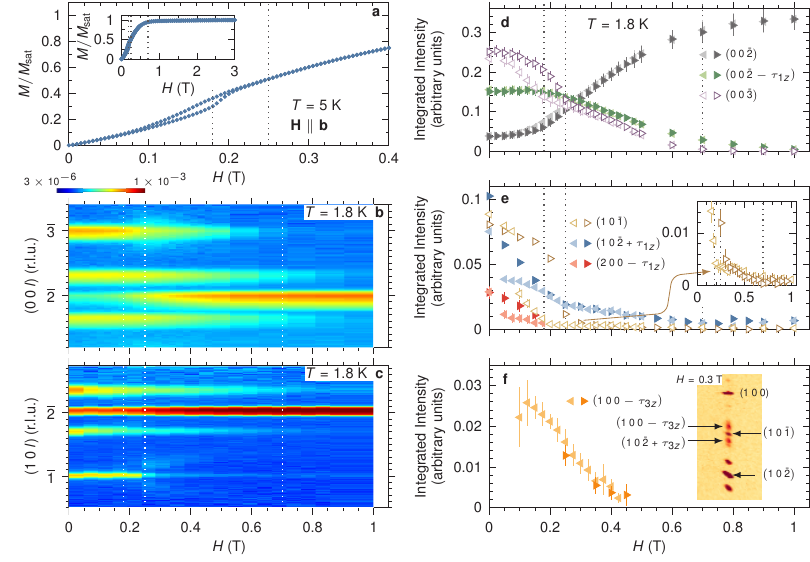}
		\caption{
			\textbf{Measured tuning of the magnetic order by an in-plane magnetic field.} \textbf{a} Magnetization versus field data for $\mathbf{H}$ parallel to the $b$ axis and a temperature of $T=5$~K. Data have been divided by the saturation value of $M_{\text{sat}}=7~\mu_{\text{B}}/\text{Eu}$ found at $H=3$~T, and the inset shows a larger field range. \textbf{b}, \textbf{c} Single-crystal $(0\,0\,l)$ (b) and $(1\,0\,l)$ (c) neutron diffraction patterns for increasing $\mathbf{H}$ parallel to the $b$ axis and $T=1.8$~K. The color scale is logarithmic and in arbitrary units. \textbf{d}--\textbf{f} The field evolution of the integrated intensities of select diffraction peaks found from fits using Gaussian lineshapes. Darker (lighter) symbols indicate increasing (decreasing) $H$. The inset to e shows data for $(1\,0\,\bar{1})$ on a different vertical scale. Vertical dotted lines in a--e indicate $H=0.18$, $0.25$, and $0.7$~T. The inset to f shows the broad weak peaks on either side of $(1\,0\,\bar{1})$ for $H=0.3$~T, which are labeled $(1\,0\,0-\tau_{3z})$ and $(1\,0\,\bar{2}+\tau_{3z})$. 
		}
		\label{Fig:Experiment}
	\end{figure*} 
	
	Figure~\ref{Fig:Experiment}a shows $M(H)$ for a temperature of $T=5$~K and $\mathbf{H}\parallel\mathbf{b}$.  With increasing field, a step begins at $H\approx0.18$~T and $M(H)$ rolls over to a shallower slope at $\approx0.7$~T. The expected saturated magnetization of $M_{\text{sat}}=7~\mu_{\text{B}}/\text{Eu}$ is reached at $\approx3$~T, where the spins are completely polarized along  $\mathbf{\hat{H}}$ \cite{Riberolles_2021}.  Upon decreasing field, hysteresis is identified between $\approx0.09$--$0.19$~T. The step and hysteresis in $M(H)$ indicate more complex behavior than the spins being continuously rotated by changing $H$. Our single-crystal neutron diffraction results, introduced next, give more insight into the field dependence of the magnetic order.
	
	We verified that the single crystal used for the neutron diffraction measurements has broken-helix ordering for $T=1.8$~K and $H=0$~T. Supplementary Figure~S$1$ of the Supplementary Information (SI) shows a comparison of the single-crystal refinement results and the measured data, where the refinement found an ordered magnetic moment per Eu of $\mu=7.0(4)~\mu_{\text{B}}$ laying in the $ab$ plane and  $\phi_{\text{rb}}=120(3)$\degree. These values are in excellent agreement with our past results \cite{Riberolles_2021}.
	
	Figures~\ref{Fig:Experiment}b and \ref{Fig:Experiment}c display $(0\,0\,l)$ and $(1\,0\,l)$ neutron diffraction patterns, respectively, for increasing  $\mathbf{H}\parallel\mathbf{b}$ and $T=1.8$~K.  Magnetic-Bragg peaks such as $(0\,0\,\bar{2}\mp\tau_{1z})$ and $(0\,0\,\bar{3})$ which correspond to the two AFM propagation vectors $\bm{\tau_1}$ and $\bm{\tau_2}=(0,0,1)$, respectively, characterize the broken-helix order and weaken with increasing field, disappearing at different field values. For $T\leq T_{\text{N}}$, $\tau_{1z}=0.323(8)\approx1/3$. The $(0\,0\,\bar{2})$ peak is an example of a purely structural-Bragg peak for $H=0$~T, and with increasing field a magnetic-Bragg peak appears on top of it, increasing in intensity with increasing field due to the spins becoming progressively polarized along $\mathbf{\hat{H}}$. Additional diffraction patterns for increasing and decreasing fields are shown in Supplementary Figs.~S$2$--S$4$.
	
	The increasing and decreasing field dependencies of the integrated intensities of select Bragg peaks are shown in Figs.~\ref{Fig:Experiment}d and \ref{Fig:Experiment}e for $T=1.8$~K. We found no remarkable changes to the full-width at half-maximum or center of any of the Bragg peaks with changing field. The apparent broadening of the $(0\,0\,\bar{3})$ Bragg peak in Fig.~\ref{Fig:Experiment}b between $H=0.25$ and $0.4$~T corresponds to the very weak and broad diffraction peaks on either side of $(1\,0\,\bar{1})$ shown in Fig.~\ref{Fig:Experiment}c and the inset to Fig.~\ref{Fig:Experiment}f. These peaks are better resolved in the  $(1\,0\,l)$ pattern due to the instrumental resolution, and the integrated intensity versus field for one of the peaks is shown in Fig.~\ref{Fig:Experiment}f. The peaks exist over a truncated field range, change position with field, and are $\approx4$ times broader than the Bragg peaks. This points to these broad peaks arising from short-range correlations, and their presence does not affect our conclusions.
	
	From Figs.~\ref{Fig:Experiment}d and \ref{Fig:Experiment}e, we identified $H=0.18$, $0.25$, and $0.7$~T as values at which the field-dependence of the integrated intensities of magnetic-Bragg peaks exhibit changes in slope for increasing $H$; a change in slope at $\approx0.18$~T is apparent in all of the datasets whereas changes in slope at $\approx0.25$~T and $\approx0.7$~T occur (within the uncertainty) for $(1\,0\,\bar{1})$, $(1\,0\,\bar{2}+\tau_{1z})$, and $(0\,0\,\bar{3})$. Dotted lines in Figs.~\ref{Fig:Experiment}a--\ref{Fig:Experiment}e indicate $H=0.18$, $0.25$, and $0.7$~T. These field values are consistent with those identified in recent reports \cite{Gen_2025,Takeda_2024,Singh_2025}.
	
	The field-dependent data for $(0\,0\,\bar{2})$ and $(0\,0\,\bar{2}-\tau_{1z})$ in Fig.~\ref{Fig:Experiment}d qualitatively mirror each other, with the $(0\,0\,\bar{2}-\tau_{1z})$ data tending to zero and the  $(0\,0\,\bar{2})$ data flattening out for $H\gtrsim1$~T, consistent with the $M(H)$ data approaching saturation. The $(2\,0\,0-\tau_{1z})$ peak disappears above $\approx0.18$~T. But, the peak is expected to be weak because $\bm{\mu}\perp\mathbf{c}$ and neutron scattering measures the component of $\bm{\mu}$ perpendicular to the scattering vector \cite{Shirane_2002}. In Fig.~\ref{Fig:Experiment}e, field hysteresis is seen for both the $(2\,0\,0-\tau_{1z})$ and $(1\,0\,\bar{2}-\tau_{1z})$ peaks from $0$~T up to $\approx0.18$~T, but up to $\approx0.25$~T for both $(0\,0\,\bar{3})$ and $(0\,0\,\bar{2}-\tau_{1z})$. Thus, some of the magnetic-Bragg peaks exhibit hysteresis over a larger field region than the $M(H)$ data. Additionally, not all of the changes in slope of the integrated intensity versus field data are apparent in $M(H)$. This points to the neutron diffraction data likely being more sensitive to changes to the magnetic ordering within individual domains.
	
	\begin{figure*}
		\centering
		\includegraphics[width=1\linewidth]{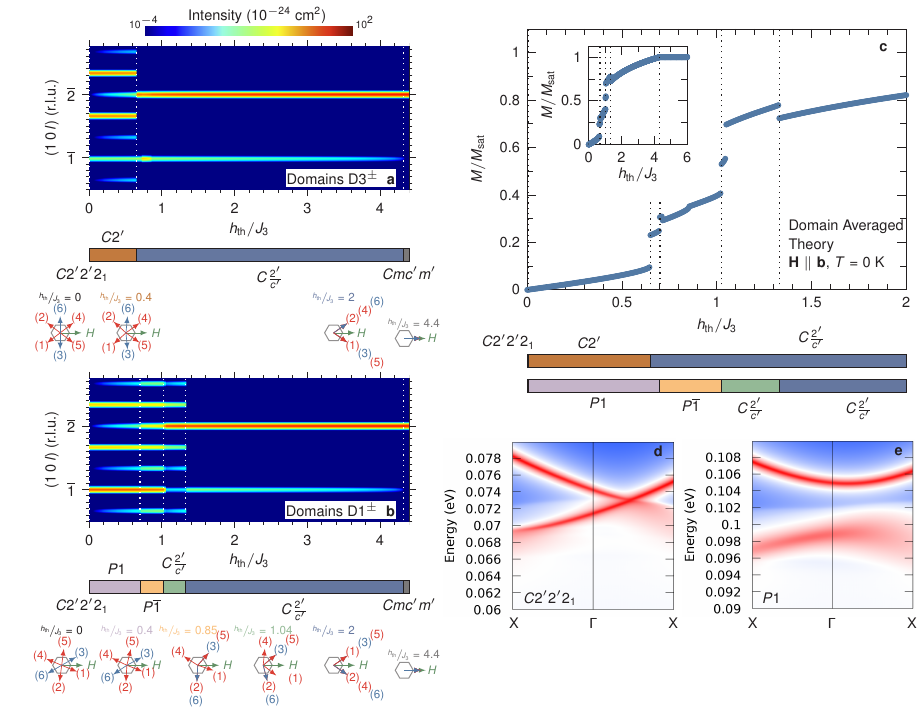}  
		\caption{
			\textbf{Calculated tuning of the magnetic order by an in-plane magnetic field.} Data are for zero temperature. \textbf{a}, \textbf{b} $(1\,0\,l)$ calculated magnetic neutron diffraction patterns for a magnetic field applied along the $b$ axis for domains D$3^\pm$ (a) and D$1^\pm$ (b). $h_{\text{th}}=g\mu_{\text{B}}H$, where $g$ is the spectroscopic splitting factor and $H$ is magnetic field. $J_3$ is the third-neighbor exchange along the $c$ axis. Dotted lines show phase transitions and bar graphs show the magnetic space groups. An example magnetic structure for each phase is also given with Eu layers numbered as in Fig.~\ref{Fig:Structure}. Diffraction intensities correspond to $1~\mu_{\text{B}}$ per Eu for a magnetic unit cell tripled along the $c$ axis. \textbf{c} The domain-averaged calculated magnetization along the $b$ axis divided by the saturated magnetization. Dotted lines show the transitions from a and b, and bar graphs similar to those in a and b are shown. \textbf{d}, \textbf{e} Electronic bands for the $[1\,1\,0]$ surface of domain D$1^+$ from density functional theory for the broken-helix (d) and $P1$ distorted-broken-helix (e) phases. 
		}
		\label{Fig:Theory}
	\end{figure*} 
	
	Neutron diffraction measurements on bulk samples typically provide information about domain-averaged order \cite{Lynn_1994,Shirane_2002}. However, the results of magnetic structure factor calculations for the broken-helix order presented in Supplementary Fig.~S$5$  show that the six magnetic domains do not equally contribute to every magnetic-Bragg peak. For example, whereas $(0\,0\,l)$ magnetic-Bragg peaks have equal contributions from each domain because $\bm{\mu}\perp\mathbf{c}$, this is not the case for $(1\,0\,l)$ magnetic-Bragg peaks. This point is further addressed below.
	
	\subsection{Analytical Theory}
	To explain our neutron-diffraction data, we assumed $\tau_{1z}=1/3$ and calculated how the ordering of each domain changes with  $\mathbf{H}\parallel\mathbf{b}$ by designing a symmetry-constrained model using the XY Hamiltonian given in Eq.~\eqref{Eq:Ham} of the Methods section which reproduces the $H=0$ broken-helix order with the value of $\phi_\text{rb}=127\degree$ previously reported \cite{Riberolles_2021}. We determined the magnetic structure of each domain at $T=0$~K for an increasing field $h_{\text{th}}=g\mu_{\text{B}}H$, where $g$ is the spectroscopic splitting factor, in step sizes of $h_{\text{th}}/J_3=0.005$ by finding the local minimum of Eq.~\eqref{Eq:Ham} at each step. We then calculated the corresponding field-dependent magnetic neutron diffraction patterns. We assumed that the domains remain equally populated and that the size of $\mu$ does not change with field.
	
	The calculated $(1\,0\,l)$ magnetic neutron diffraction patterns for D$3^+$ and D$1^+$ are shown in Figs.~\ref{Fig:Theory}a and \ref{Fig:Theory}b, respectively. The corresponding negative chirality domains have the same diffraction patterns because the patterns do not depend on chirality. Additionally, the patterns for D$1^+$ and D$2^+$ are the same due to their spins' similar initial orientations with respect to $\mathbf{\hat{H}}$. The extent of each magnetic phase is indicated by the bar graphs and dotted white lines. The MSG and example spin orientations for each phase are given. 
	
	For high field, all of the domains eventually order into the canted-A-type structure (MSG $C2^\prime/c^\prime$) before reaching the field-polarized phase ($Cmc^\prime m^\prime$). For lower field, however, D$3^\pm$ have canted broken-helix ($C2^\prime$) structures, whereas D$1^\pm$ and D$2^\pm$ have distorted-broken-helix ($P1$), canted-orthogonal ($P\bar{1}$), or two-angle-canted ($C2^\prime/c^\prime$) orders. The two-angle-canted order resembles a magnetic fan and, unlike the canted-A-type order, has a tripled unit cell along $c$.
	
	Data in Fig.~\ref{Fig:Experiment} can be semi-quantitatively compared with the calculations in Fig.~\ref{Fig:Theory}.   However, we expect phase transitions to appear less sharp in the measured data due to the finite temperature. In addition, the measured diffraction patterns include structural-Bragg peaks. Magnetic-Bragg peaks $\approx4$ orders of magnitude weaker than the others are also present in Figs.~\ref{Fig:Theory}a and \ref{Fig:Theory}b at $(0,0,2/3)$ positions, but are not seen in Figs.~\ref{Fig:Experiment}b and \ref{Fig:Experiment}c. These peaks are likely too weak to have been detected by the measurements, and whether they exist does not affect our conclusions. 
	
	Considering the steps in the calculated $M(h_\text{th}/J_3)$ data in Fig.~\ref{Fig:Theory}c, the step at $H\approx0.18$~T in the measured $M(H)$ data in Fig.~\ref{Fig:Experiment}a for increasing field, the hysteresis in the neutron diffraction and magnetization data, and the domain-dependence of the intensity of certain Bragg peaks, we infer that the values of $H=0.18$, $0.25$, and $0.7$~T previously identified approximately correspond to field values at which changes to magnetic ordering within certain domains occurs. In particular, we deduce that domains D$3^\pm$ enter the canted-A-type phase above $\approx0.18$~T based on the calculated intensity of the $(1\,0\,2+\tau_{1z})$ Bragg peak being dominated by D$3^\pm$ at low fields, that hysteresis may be expected to accompany the flipping of some of the spins in D$3^\pm$ from against to along the field direction, and that a step occurs in the calculated $M(H)$ curve when D$3^\pm$ transition to the canted-A-type phase. 
	
	Similarly, we infer that D$1^\pm$ and D$2^\pm$ transition from the distorted-broken-helix to the canted-orthogonal phase at $\approx0.25$~T, and to the two-angle-canted phase above $\approx0.7$~T. This latter point agrees with the $(1\,0\,\bar{1})$ magnetic-Bragg peak being dominated by D$1^\pm$  and D$2^\pm$ prior to transitioning to the two-angle-canted phase, as shown in Figs.~\ref{Fig:Theory}a and \ref{Fig:Theory}b. The transition from the two-angle-canted to the canted-A-type phase for D$1^\pm$ and D$2^\pm$ is not visible in the measured $M(H)$ data but corresponds to the complete disappearance of the $\bm{\tau_{1}}$-Bragg peaks in Figs.~\ref{Fig:Experiment}b and \ref{Fig:Experiment}c for $H\gtrsim1$~T. The in-plane-field evolutions for the orderings within domains that we predict are similar to the recently proposed in-plane-field induced fan-like phase which closes with field and eventually transitions to a forced-ferromagnetic state \cite{Takeda_2024, Singh_2025, Gen_2025, Soh_2023}.
	
	\section{Discussion}
	\begin{figure*}
		\centering
		\includegraphics[width=1\linewidth]{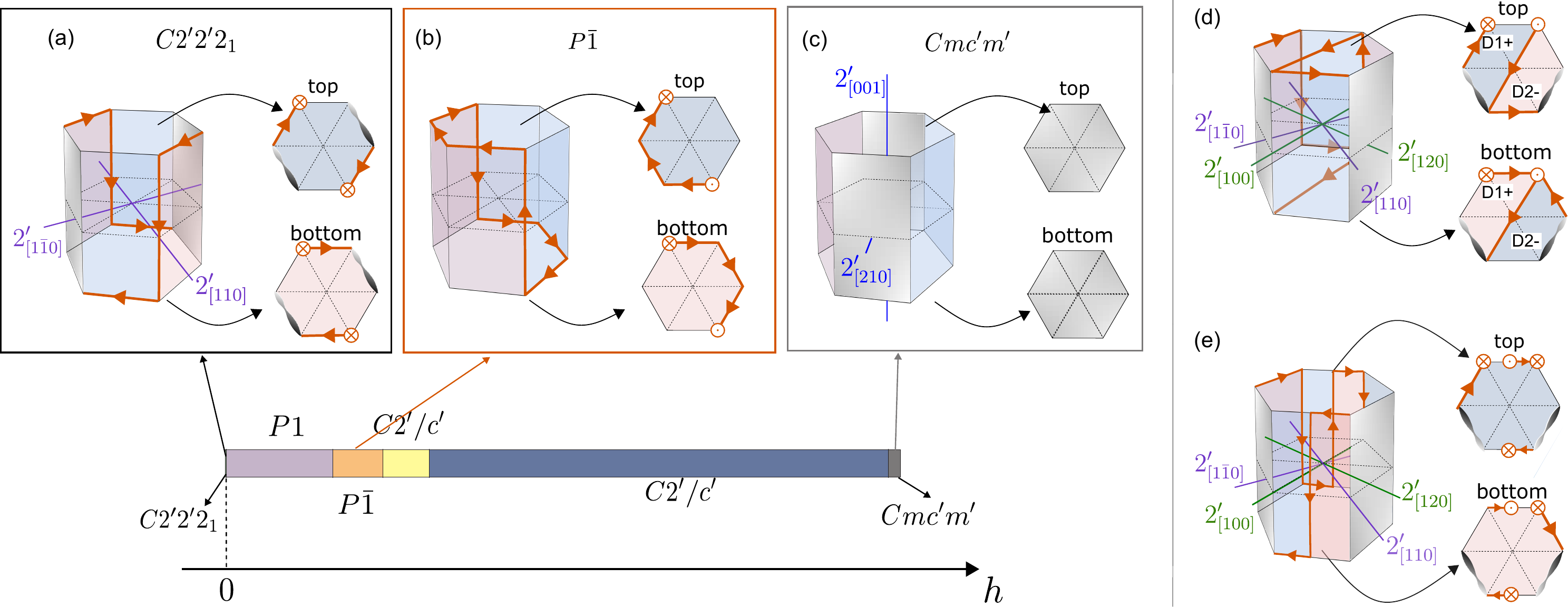}
		\caption{
			\textbf{In-plane magnetic-field evolution of surface-Dirac and hinge states for magnetic domains of EuIn$_2$As$_2$.} \textbf{a} Hinge states for domain D$1^+$ in the broken-helix magnetic ground state. Light-gray surfaces are metallic with gapless Dirac states. Red and blue surfaces have gapped surface Dirac states with effective masses $m^*<0$ and $m^*>0$, respectively. Hinge states (orange arrows) develop between adjacent surfaces with opposite signs for $m^*$. An in-plane magnetic field $\mathbf{H}$ applied parallel to the $b$ axis drives bulk magnetic phase transitions which induce topological changes in the surface Dirac states. \textbf{b} Field-induced canted-orthogonal phase for D$1^+$. The $P\bar{1}$ symmetry enforces inversion-related surfaces to have opposite signs for $m^*$. \textbf{c} Field-polarized phase for which the $Cmc^\prime m^\prime$ symmetry leads to metallic surfaces perpendicular to $\mathbf{\hat{H}}$. We show a symmetry-allowed pattern for the signs of $m^*$ where no hinge states develop. \textbf{d} The intersection of D$1^+$ and D$2^-$ at a $[1\,2\,0]$ domain wall for the broken-helix phase. Hinge states pinned to the wall occur at the intersections with the top and bottom surfaces because $m^*$ changes signs across the intersections. \textbf{e} The intersection of D$1^+$ and D$2^+$ at a $[2\,1\,0]$ domain wall for the broken-helix phase. Hinge states pinned to the domain wall occur at the side surfaces.  
		}
		\label{Fig:Hinges}
	\end{figure*}
	
	Our results imply a large region of magnetic field for which distinct magnetic domains can be tuned to have different magnetic symmetries. We analyze possible impacts this can have on the topological hinge states using the relationship between the magnetic symmetry of a domain for a given field value and the presence of gapped or gapless Dirac states on outer surfaces of the sample. Because symmetry relates the sign of $m^*$ for gapped surfaces, field-induced symmetry changes can control the hinge states.
	
	We previously reported density-functional-theory (DFT) results for EuIn$_2$As$_2$ showing that the bulk topological electronic gap predicted by DFT remains open with increasing magnetic field through entry into the field-polarized state~\cite{Riberolles_2021}. Our new DFT results in Figs.~\ref{Fig:Theory}d and \ref{Fig:Theory}e now specifically show that the gapless Dirac states for the $[1\,1\,0]$ surface in the broken-helix phase become gapped for the $P1$ distorted-broken-helix phase. This demonstrates that weak in-plane magnetic fields ($H<0.18$~T) switch certain surface Dirac states from gapless to gapped by reducing the bulk symmetry.
	
	An example hinge-state pattern for the broken-helix order is shown in Fig.~\ref{Fig:Hinges}a for domain D$1^+$.  As explained in the Introduction, surfaces normal to $2^\prime$ axes are gapless whereas other surfaces are gapped. Opposite side surfaces related by $2_{[0\,0\,1]}$ symmetry necessarily have the same signs for $m^*$, but $m^*$ has opposite signs for surfaces related by $2^\prime$ \cite{Zhang_2012, Tanaka_2020}. Since $\mathcal{P}$ relates domains D$1^-$ and D$1^+$, this demands reversed signs of $m^*$ for the surfaces of D$1^-$. Thus, the hinge-state paths for D$1^-$ and D$1^+$ have opposite directions.
	
	Next, we predicted above that for a weak in-plane field $\mathbf{H}\parallel\mathbf{b}$ the $2_{[0\,0\,1]}$ and $2^\prime$ symmetry axes are lost for D$1^\pm$ and D$2^\pm$ as the MSG evolves first to $P1$ and then to $P\bar{1}$. This change of magnetic symmetry induces a topological phase transition by opening gaps on some of the surfaces. This is seen by comparing Figs.~\ref{Fig:Hinges}a and \ref{Fig:Hinges}b, where for P$\bar{1}$ all the surfaces are gapped and opposite side surfaces have opposite signs for $m^*$. Thus, during the $C2^\prime2^\prime2_1\rightarrow P1\rightarrow P\bar{1}$ evolution,  $m^*$ for some of the side surfaces changes sign and a phase transition occurs when $m^*$ crosses $0$. The hinge-state patterns for the P$\bar{1}$ canted-orthogonal and $C2^\prime2^\prime2_1$ broken-helix phases must therefore be different. With further increasing field, $2^\prime$ symmetry axes eventually reemerge in the canted-A-type and field-polarized phases, with gapless Dirac states for surfaces perpendicular to $\mathbf{H}$. We show in Fig.~\ref{Fig:Hinges}c a possible symmetry-allowed pattern for the surface masses of D$1^+$ in the field-polarized phase which does not lead to hinge states. 
	
	Our analysis has also found that hinge states appear on magnetic domain walls intersecting outer surfaces of the crystal if the walls separate gapped surfaces with opposite signs for $m^*$. The relative sign of $m^*$ for surfaces of different domains is dictated by symmetry. For example, the sign of $m^*$ would be opposite for the two sides of a domain wall if the two magnetic states are related by $\mathcal{P}$. Therefore, a domain wall between opposite chirality states [such as shown in Fig.~\ref{Fig:Examples}(e)] carries a hinge state. Yet, this situation occurs in many other domain wall configurations as well with hinges occurring on top, bottom and side surfaces. For example, as shown in Fig.~\ref{Fig:Hinges}d for the broken-helix phase with domains D$1^+$ and D$2^-$ intersecting along a $[1\,2\,0]$ domain wall. Since top and bottom surfaces are related by $2^\prime$ (even though the direction of $2^\prime$ in the $ab$ plane differs for the two domains), hinge states pinned to the domain wall must exist where the wall intersects the top and bottom surfaces. Another example showing a hinge state on a side surface that is pinned to a domain wall between D$1^+$ and D$2^+$  is shown in Fig.~\ref{Fig:Hinges}(e). Based on our findings, one can conceive complex field-tunable hinge-state patterns.
	
	Next, we address the issue of the observed slight incommensurability of the magnetic order and explain why this does not invalidate our hinge mode analysis. Performing a group theory analysis of the possible magnetic states for other commensurate values of $\tau_z$ increasingly close to the experimental value of $0.323(8)$, we found magnetically ordered states that preserve $2^\prime$ operations, which is essential for the quantization of the $\theta$ angle and the emergence of hinge modes between adjacent gapped surfaces related by  $2^\prime$. Therefore, different MSGs with $2^\prime$ symmetries can yield different patterns of the hinge modes, but do not invalidate their existence. Furthermore, as mentioned in the Introduction, we expect that the incommensurability can be treated perturbatively, as the magnetic state is interrupted only every $100$th Eu layer. Whereas this implies that surfaces normal to $2^\prime$ acquire a gap, we expect that the sign of $m^*$ on already gapped surfaces identified in the $\tau_z=1/3$ case remains approximately robust.
	
	The breaking of $2^\prime$ symmetry by incommensurate magnetic order with $\tau_z \approx 1/3$ opens a gap on surfaces normal to the nominal $2^{\prime}$ axes, which are gapless in the commensurate case. For surfaces that are already gapped for commensurate order with surface mass $m^*$, we expect the sign of $m^*$ remains unchanged provided the incommensurability length scale $l_\text{incomm}$, defined as the distance over which the magnetic modulation along $z$ accumulates a $\pi$ phase shift relative to commensurate order, is sufficiently large. For top and bottom surfaces, $l_{\text{incomm}}$ should be compared with the bulk gap length scale $l_\text{bulk} \propto 1/\Delta_{bulk}$, which characterizes variation of the localized wavefunction along $z$. For the side surfaces, this variation is characterized by the scale $l_{\text{surface}} \propto 1/m^*$. Thus, as long as $l_{\text{incomm}} \gg l_{\text{bulk}}, l_{\text{surface}}$, the hinge-state pattern is expected to remain robust against incommensurability.
	
	Regarding our sample not being a bulk insulator, quantized conductivity for the hinge modes is expected to be lost in the presence of bulk metallic states. However, hinge modes may contribute significantly to surface transport due to their robustness to disorder and inhomogeneities, even though quantization may be lost. Furthermore, Ca substitution for Eu is a potential viable route towards an insulating bulk \cite{Yan_2024} and may be used to test the robustness of the hinge modes.
	
	Notably, noncollinear magnetic order, such as in EuIn$_2$As$_2$, generally enables a rich functionality for controlling the location of hinge states by a relatively weak magnetic field. Such a situation can occur at an intersection of three domains, as illustrated in  Figs.~\ref{Fig:Examples}e--\ref{Fig:Examples}g, where the field direction controls the sign of the emergent $m^*$ and tunes the locations of hinge-states. Hinge states pinned to domain walls will persist with changing $\mathbf{H}$ as long as $m^*$ for the surfaces intersecting the wall and supporting the hinge states do not switch sign. This means that $\mathbf{H}$ is a mechanism for continuously moving domain walls and pinned hinge states to another position in the sample, creating, for example, functionality such as a topological electric switch.
	
	Finally, the extent to which a hinge state penetrates into a domain wall needs further clarification. Theoretically, it is set by the inverse of $m^*$ and can thus be calculated {\it ab initio}. However, calculations for an AXI show that the first few layers from the surface contribute to the QAH-type conductivity~\cite{Varnava_2018} whereas domains walls of an AFM are known to be $\approx 10$--$100$~nm thick. Future work aimed at directly investigating hinge states pinned to domain walls should involve direct probes that can differentiate between bulk and surface states.
	
	\section{Conclusion}
	
	We have shown how noncollinear magnetic ordering of an AXI can be manipulated by an in-plane magnetic field to tune magnetic symmetry and, thus, the topologically protected surface states. Our measurements and theoretical analysis describe how magnetic domains in EuIn$_2$As$_2$ evolve from the broken-helix ground state to the field-polarized phase for $\mathbf{H}\parallel\mathbf{b}$, which, in turn, has allowed us to demonstrate how complex hinge-state patterns arising in the AXI phase are tuned by symmetry changes to each domain. We find a further degree of topological tunability in the existence of hinge states pinned to magnetic walls, where the associated chiral-conduction pathways can move with the domain wall.
	
	Control of topological properties is at the cutting edge of contemporary materials physics research \cite{Qiu_2023}, as topological quantum materials host remarkable properties such as dissipationless chiral-charge transport and platforms to study exotic quaisparticles and interactions \cite{Qiu_2025}.  Our results show multiple degrees of robust tuning of topological electronic states by a weak magnetic field in a TI with noncollinear magnetic order and highlight the complex intertwining of symmetry and topology.
	
	\section{Methods} \label{Methods}
	\subsection{Magnetization}
	Single crystals of EuIn$_2$As$_2$ were flux grown as described previously \cite{Riberolles_2021}. Measurements of the magnetization $M$ were made at a temperature of $T=1.8$~K and in a magnetic field up to $H=5$~T utilizing a Quantum Design, Inc., Magnetic Property Measurement System with a superconducting quantum interference device. The field was applied along the $\mathbf {b}$ crystalline axis after the samples were oriented using an X-ray Laue camera.
	\subsection{Neutron Diffraction}		
	Single-crystal neutron diffraction was carried out at the Spallation  Neutron Source, Oak Ridge National Laboratory, USA, using the CORELLI time-of-flight (TOF) spectrometer \cite{Ye_2018}. CORELLI utilizes a pulsed white neutron beam and TOF analysis of the neutrons scattered into a large $2$D position sensitive detector to generate diffraction patterns spanning large ranges of neutron-momentum transfer.
	
	A flat-plate single-crystal with a mass of $18.4$~mg was glued to an Al pin and aligned with the $(h\,0\,l)$ reciprocal lattice plane horizontal. The sample was cooled down to $T=1.8$~K inside of a vertical-field superconducting magnet such that $\mathbf {H}\parallel\mathbf {b}$. Data were collected at discrete fields upon ramping the field up or down. The sample was warmed up to $20$~K between field cycles and then cooled in zero field in order to start each cycle from a virgin state. The sample was rotated about its vertical axis and measured for various crystal orientations. Data were normalized to scattering from vanadium to account for the non-linear wavelength distribution in the pulsed beam.
	
	Data analysis, including corrections for the large thermal neutron absorption of Eu were performed using \textsc{mantid}~\cite{Mantid} and \textsc{mag$2$pol} \cite{mag2pol}. Symmetry analysis of the magnetic phases was performed using the Bilbao Crystallographic Server~\cite{bilbao} and custom software. Some figures include diagrams made using \textsc{vesta} \cite{Momma_2011}.
	
	\subsection{Analytical Theory}
	Insight into the field dependence of the different magnetic-Bragg peaks has been gained by deriving a symmetry-constrained $J_0$-$J_1$-$J_2$-$J_3$-$K_1$ XY Hamiltonian with nearest ($J_1$), next-nearest ($J_2$),  third-neighbor ($J_3$) effective exchange interactions, and a nearest-neighbor biquadratic interaction ($K_1$) along the $\mathbf {c}$ axis \cite{Pimpinelli_1989} that reproduces the $H=0$ broken-helix order. We represent each Eu layer by a fixed-length classical spin $\mathbf {S}$ at position $i$ or $j$ along $\mathbf{c}$ and orient the spin within the $\mathbf {ab}$ plane by an implied large in-plane magnetic anisotropy. The antiparallel orientation of spins between third-neighbor layers in the broken helix suggests $J_3$ is significant and AFM ($J_3>0$).  We also introduce a nearest-neighbor biquadratic exchange coupling $K_{1}>0$ \cite{Anderson_1963,Hoffmann_2020} to stabilize the broken-helix from the manifold of degenerate states, including a regular helix.
	
	The in-field Hamiltonian is 
	\begin{equation}
		\begin{split}
			\begin{aligned}
				\mathcal{H}&=\mathcal{H}_{\text{in-plane}}+J_1\sum_{\langle ij \rangle}   \mathbf {S}_i \cdot \mathbf {S}_j+J_2\sum_{\langle\langle ij \rangle\rangle}\mathbf {S}_i \cdot \mathbf {S}_j\\&+J_3\sum_{\langle\langle\langle ij \rangle\rangle\rangle}\mathbf {S}_i \cdot\mathbf {S}_j
				+K_{1}\sum_{\langle ij \rangle}\left(S^x_iS^x_j+S^y_iS^y_j\right)^2\\
				&+\frac{D_{xy}}{2}\sum_i\left[\left(S_i^x+iS_i^y\right)^6+\left(S_i^x-iS_i^y\right)^6\right]\\&-g\mu_{\text{B}}\sum_i\mathbf {S}_i\cdot \mathbf {H}\ ,\label{Eq:Ham}
			\end{aligned}
		\end{split}
	\end{equation}
	where $g$ is the spectroscopic splitting factor, $\mathcal{H}_{\text{in-plane}}$ accounts for the effective ferromagnetic interaction ($J_0$) within each Eu layer and the large in-plane anisotropy, and $D_{xy}$ is a six-fold magnetic anisotropy constant pinning on layer of spins (the blue spins) in Fig.~$2$ of the main text along the specific in-plane axis identified by experiment \cite{Riberolles_2021}. Using a variational method and $S_i=1$, we found that $J_1=-0.5$, $J_2=-0.6$, $K_{1}=0.2$, $D_{xy}=-0.02$, and $J_3=1$ recreates the $H=0$ broken-helix order with the value of $\phi_\text{rb}=127\degree$ previously reported \cite{Riberolles_2021} while using an extremely small value for $D_{xy}$. Further details of the ground-state phase diagram and the evolution of the broken-helix in an in-plane field are given in the SI (Supplementary Figs.~S$6$ and S$7$) and Ref.~\cite{nedic2023effects}.
	
	\subsection{Density Function Theory Calculations}
	Band structures using density functional theory (DFT) with spin-orbit coupling were calculated with the Perdew-Burke-Ernzerhof (PBE) exchange-correlation functional, a plane-wave basis set, and the projected-augmented-wave method as implemented in \textsc{vasp}~\cite{Kresse_1996,Kresse_1996_2}. To account for the half-filled strongly localized Eu $4f$ orbitals, a Hubbard-like~\cite{Dudarev_1998} $U$ value of $5.0$~eV is used. For different helical magnetic structure with $\bm{\tau}\approx(0,0,1/3)$, i.e.\ the hexagonal unit cell is tripled along the $\mathbf {c}$ axis with atoms fixed in their bulk positions. A Monkhorst-Pack $12\times12\times1$ $k$-point mesh with a Gaussian smearing of $0.05$~eV, including the $\Gamma$ point, and a kinetic-energy cutoff of $250$~eV have been used. To search for possible band gap closing points in the full BZ, a tight-binding model based on maximally localized Wannier functions~\cite{Marzari_2012} was constructed to reproduce closely the bulk band structure, including spin-orbit coupling in the range of $E_\text{F} \pm1$~eV with Eu $sdf$, In $sp$ and As $p$ orbitals as implemented in \textsc{wanniertools}~\cite{Wu_2017}. 
	
	\section{Data availability}	
	Datasets will be made publicly available once the manuscript is accepted for publication.
	
	\section{Code availability}	
	The code used for this study may be made available to qualified researchers upon a reasonable request.
	
	\section{Acknowledgments}	
	Sample synthesis, neutron diffraction experiments, electronic structure calculations, symmetry analysis, and analytical calculations were supported by the Center for the Advancement of Topological Semimetals (CATS), an Energy Frontier Research Center funded by the  U.S.\ Department of Energy (DOE) Office of Science (SC), Office of Basic Energy Sciences (BES), through the Ames National Laboratory under Contract No.\ DE-AC$02$-$07$CH$11358$. Magnetization measurements were made at the Ames National Laboratory through the Complex States, Emergent Phenomena, \& Superconductivity in Intermetallic \& Metal-like Compounds Field Work Proposal supported by the U.S.\ DOE SC, BES, Division of Materials Sciences and Engineering. Ames National Laboratory is operated for the U.S.\ DOE by Iowa State University under Contract No.\ DE-AC$02$-$07$CH$11358$. A portion of this research used resources at the Spallation Neutron Source, which is a U.S.\ DOE SC User Facility operated by Oak Ridge National Laboratory. The beam time was allocated to Corelli on proposal number IPTS-$26414$.
	
	
	\section{Author contributions}
	SXMR, FY, RJM, and BGU performed the neutron diffraction experiments and analyzed the results. BK, SLB, and PCC synthesized the samples and performed magnetization measurements. LLW performed the electronic structure calculations. AMN, TVT, JA, VLQ, and PPO performed the symmetry analysis and analytical calculations. SXMR, AMN, TVT, VLQ, RJM, PPO, and BGU wrote the manuscript with input from all of the authors.		
	
	\section{Competing interests}
	All authors declare no financial or non-financial competing interests. 	
	

	\newpage
	
	\makeatletter
	\setcounter{equation}{0}
	\setcounter{figure}{0}
	\setcounter{table}{0}
	\setcounter{section}{0}
	
	\renewcommand{\theequation}{S\arabic{equation}}
	\renewcommand{\thefigure}{S\arabic{figure}}
	\renewcommand{\thetable}{\S\arabic{table}}
	
	\renewcommand{\figurename}{\textbf{Supplementary Fig.}}
	\renewcommand{\tablename}{\textbf{Supplementary Table}}
	\renewcommand{\bibnumfmt}[1]{[S#1]}
	\renewcommand{\citenumfont}[1]{S#1}
	
	\begin{center}
		\textbf{Symmetry tuning topological states of an axion insulator with noncollinear magnetic order \\ \normalfont (Supplementary Information)}
	\end{center}
	\vspace{2ex}

	\section*{Supplementary Note 1}
	\subsection*{Further neutron diffraction results}
	\begin{figure}
		\includegraphics[width=1.0\linewidth]{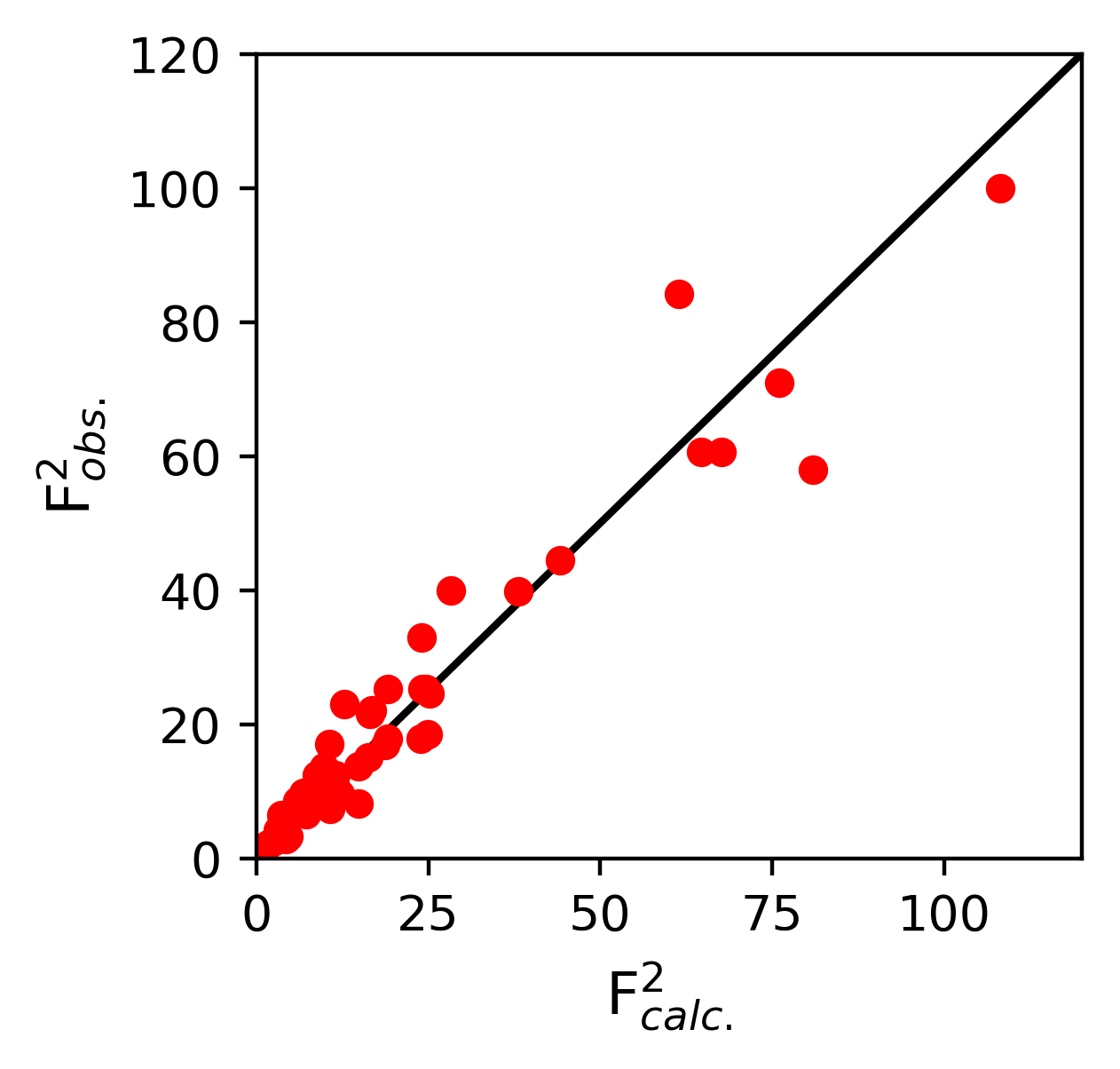}
		\caption{ \textbf{Observed (obs.) versus calculated (calc.) values of the sum of the squares of the structural and magnetc neutron structure factors for broken-helix ordering of EuIn$_{\bm{2}}$As$_{\bm{2}}$.} The observed data are for a temperature of $T=1.8$~K and zero magnetic field.}
		\label{Fig_SI:Refine}
	\end{figure} 
	
	Supplementary Figure~\ref{Fig_SI:Refine} shows results from our single-crystal refinement to the magnetic and nuclear Bragg peak integrated intensities determined from neutron diffraction measurements made at a temperature of $T=1.8$~K and a magnetic field of $H=0$~T,  and after being adjusted to account for neutron absorption \cite{mag2pol}. We used the crystallographic information from Ref.~\cite{Goforth_2008}. The final refinement had reasonable goodness-of-fit values for both the nuclear ($R_F=10.4$) and magnetic phases ($R_F=11.2$), and determined broken-helix magnetic order with an ordered Eu magnetic moment of $\mu=7.0(4)~\mu_\text{B}$ and a helix angle of $\phi_{\text{rb}}=120(3)$\degree. These values are both consistent with those previously reported \cite{Riberolles_2021}. The slight differences in $\mu$ and $\phi_{\text{rb}}$ between the previously reported values may be due to sample-dependent effects, such as localized strain \cite{Donoway_2023}, or differences in measurement temperature.

	\begin{figure}
		\includegraphics[width=1\linewidth]{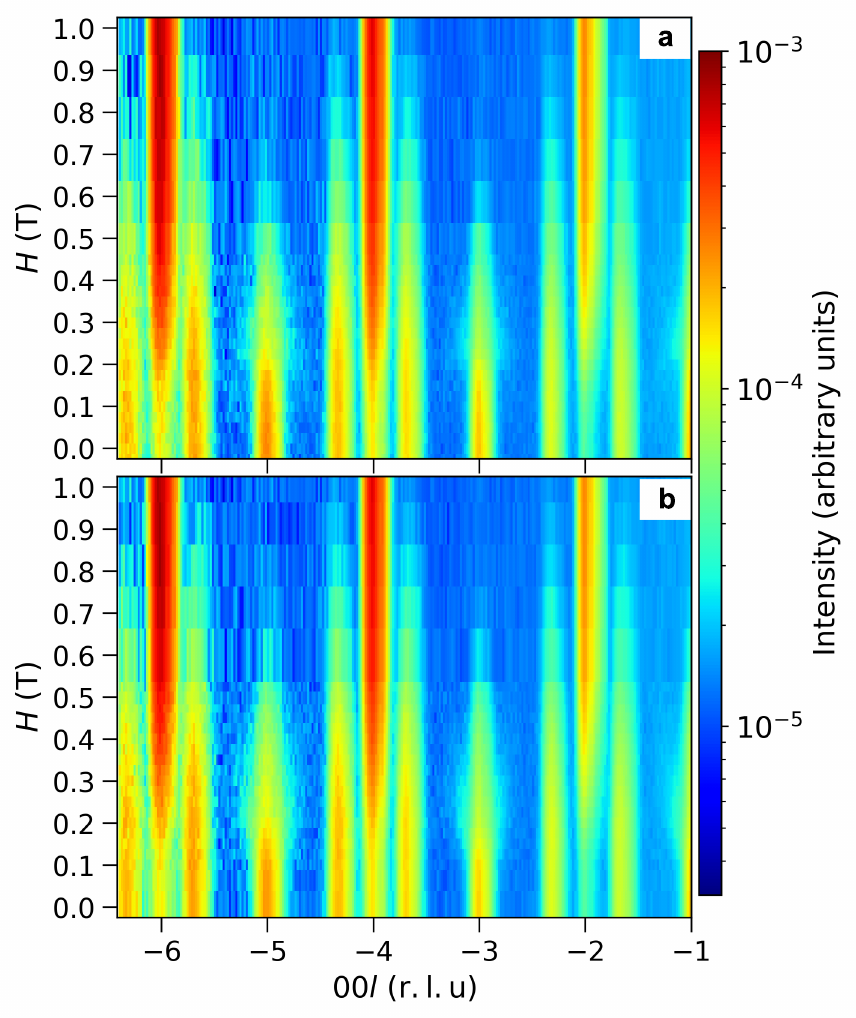}
		\caption{  \textbf{EuIn$_{\bm{2}}$As$_{\bm{2}}$ single-crystal neutron-diffraction patterns for $\bm{(0\,0\,l)}$ as a function of magnetic field for $\bm{T=1.8}$~K.} Data in \textbf{a} were collected while ramping up the field (after zero-field cooling), and data in \textbf{b} were taken subsequently, while ramping the field down.}
		\label{Fig_SI:00l}
	\end{figure} 
	
	\begin{figure}
		\includegraphics[width=1\linewidth]{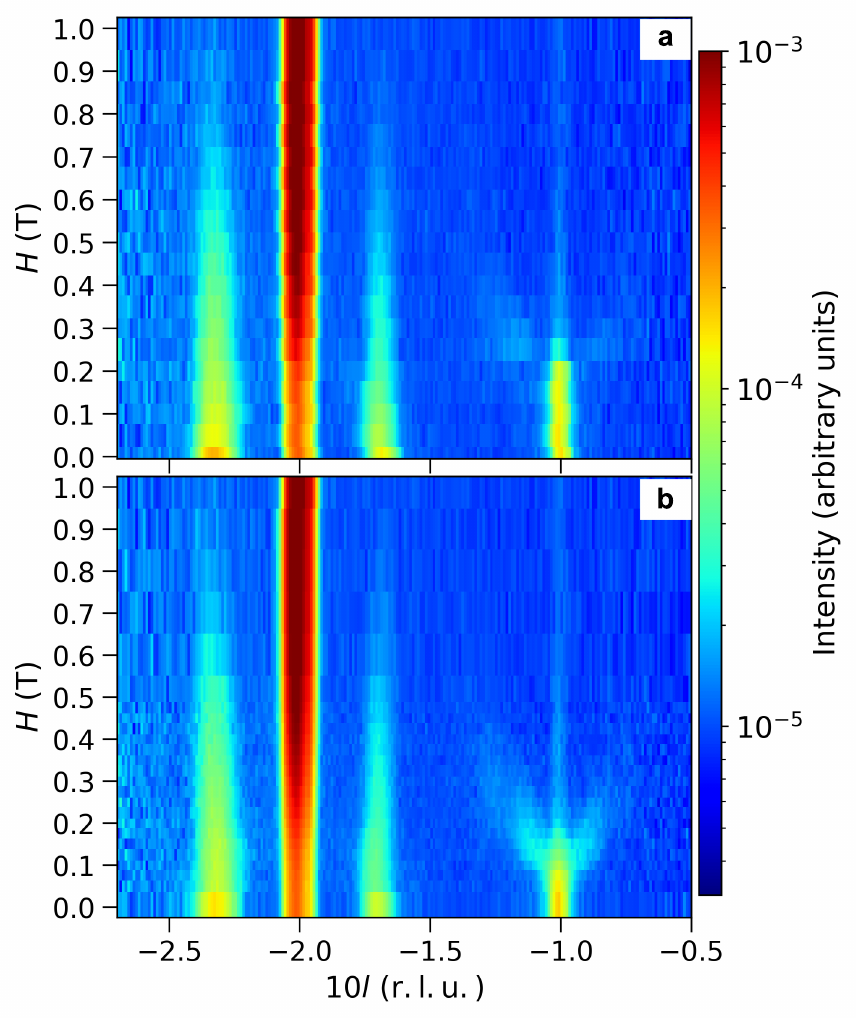}
		\caption{  \textbf{EuIn$_{\bm{2}}$As$_{\bm{2}}$ single-crystal neutron-diffraction patterns for $\bm{(1\,0\,l)}$ as a function of magnetic field for $\bm{T=1.8}$~K.} Data in \textbf{a} were collected while ramping up the field (after zero-field cooling), and data in \textbf{b} were taken subsequently, while ramping the field down.}
		\label{Fig_SI:10l}
	\end{figure} 
	
	\begin{figure}
		\includegraphics[width=1\linewidth]{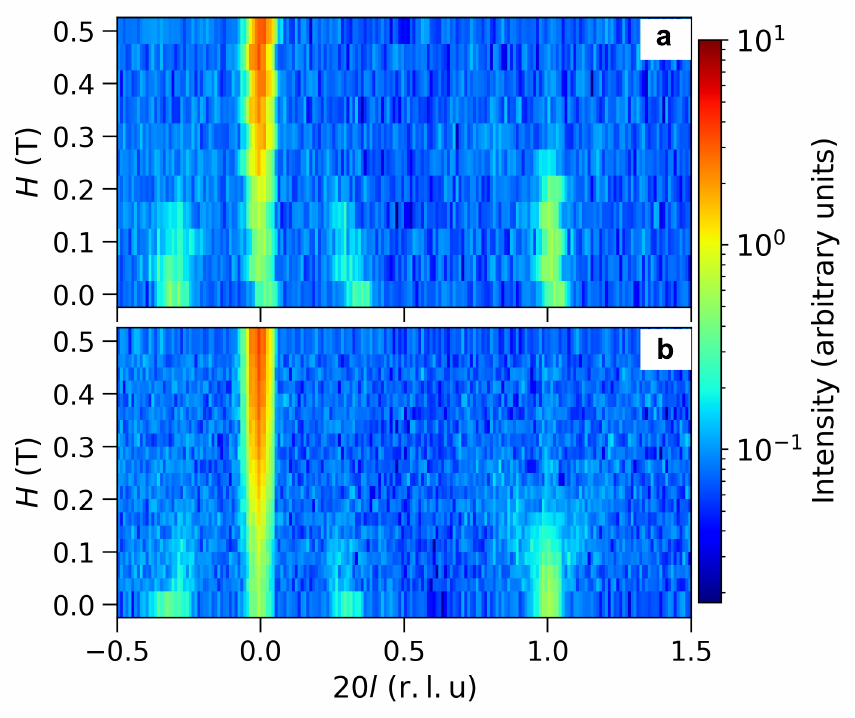}
		\caption{  \textbf{EuIn$_{\bm{2}}$As$_{\bm{2}}$ single-crystal neutron-diffraction patterns for $\bm{(2\,0\,l)}$ as a function of magnetic field for $\bm{T=1.8}$~K.} Data in \textbf{a} were collected while ramping up the field (after zero-field cooling) and data in \textbf{b} were taken subsequently, while ramping the field down. }
		\label{Fig_SI:20l}
	\end{figure} 
	
	Supplementary Figures~\ref{Fig_SI:00l}a,  \ref{Fig_SI:10l}a, and \ref{Fig_SI:20l}a display the $(0\,0\,l)$, $(1\,0\,l)$, and $(2\,0\,l)$, respectively,  field-dependent neutron diffraction patterns for $T = 1.8$~K and an increasing magnetic field applied parallel to the $b$ axis. Similar patterns for subsequent decreasing field measurements are shown in Supplementary Figs.~\ref{Fig_SI:00l}b, \ref{Fig_SI:10l}b, and \ref{Fig_SI:20l}b.

	\begin{figure}
		\includegraphics[width=1\linewidth]{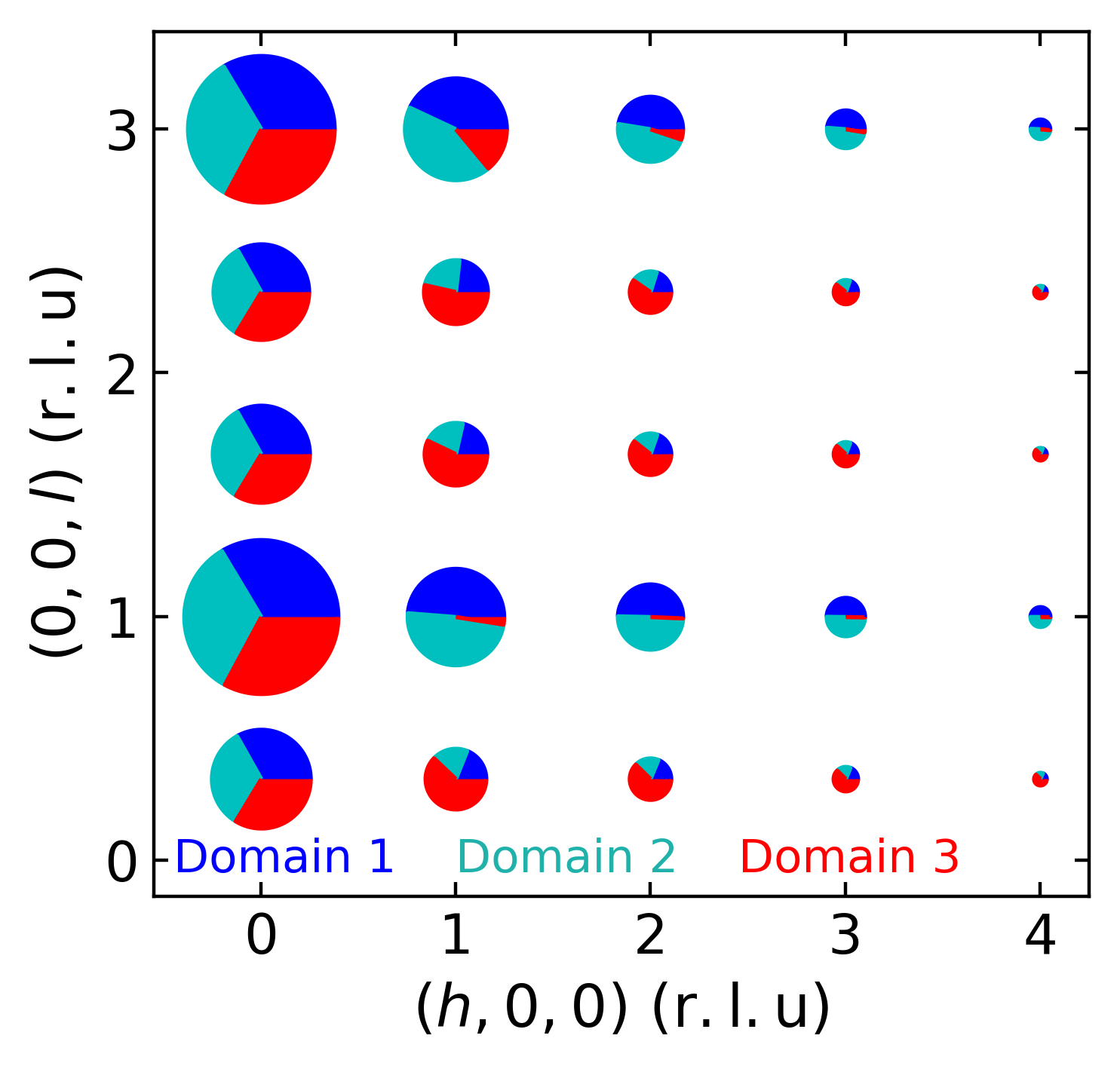}
		\caption{ \textbf{Domain-resolved calculated $\bm{(h\,0\,l)}$ magnetic diffraction intensities for the zero-field broken-helix ground state.} The pie charts display both the individual magnetic domain contributions (colors), and the relative total intensities (circle sizes) of each peak.}
		\label{Fig_SI:Dom_Contr}
	\end{figure}
	
	Supplementary Fig.~\ref{Fig_SI:Dom_Contr} displays the calculated individual magnetic domain contributions to $(h\,0\,l)$ magnetic-Bragg peaks from the broken-helix order. As detailed in the main text, the zero-field ground state has magnetic space group (MSG) $C2^\prime2^\prime2_1$ and orthorhombic symmetry. This consequently forms six magnetic domains when ordering on the EuIn$_2$As$_2$ hexagonal chemical lattice; three domains related by $120\degree$ rotations and each of the three domains can have $+$ or $-$ chirality. In this section, we refer to only three domains and do not differentiate between positive or negative chirality, since they were indistinguishable by our measurements.
	
	Supplementary Fig.~\ref{Fig_SI:Dom_Contr} shows that $(0\,0\,l)$ Bragg reflections contain equal contributions from each domain because the ordered magnetic moment lies in the $ab$-plane and neutron diffraction measures the ordered moment perpendicular to the direction of the momentum transfer. Other reflections do not have equal intensity for each domain, which grants some sensitivity to domain-related features: $(h\,0\,l)$ magnetic-Bragg reflections with $h\neq0$ display unequal domain contributions; $\bm{\tau_1}=(0,0,\tau_{1z})$, $\tau_{1z}\approx\frac{1}{3}$, magnetic-Bragg peaks are dominated by domain $3$ (D$3$) and $\bm{\tau_2}=(0,0,1)$ magnetic-Bragg peaks are dominated by D$1$ and D$2$. Noticeably, the $(1\,0\,1)$ magnetic-Bragg peak is dominated at $98$\% by the combined contributions of D$1$ and D$2$, whereas reflection $(1\,0\,2-\tau_{1z})$ is dominated at $53$\% by the contribution of D$3$.
	
	\section*{Supplementary Note 2}
	\subsection*{Analytical model for the field dependence of the broken helix}
	
	The in-field Hamiltonian reproduced from Eq.~($1$) of the main text is
	\begin{equation}
		\begin{split}
			\begin{aligned}
				\mathcal{H}&=\mathcal{H}_{\text{in-plane}}+J_1\sum_{\langle ij \rangle}   \mathbf {S}_i \cdot \mathbf {S}_j+J_2\sum_{\langle\langle ij \rangle\rangle}\mathbf {S}_i \cdot \mathbf {S}_j\\&+J_3\sum_{\langle\langle\langle ij \rangle\rangle\rangle}\mathbf {S}_i \cdot\mathbf {S}_j
				+K_{1}\sum_{\langle ij \rangle}\left(S^x_iS^x_j+S^y_iS^y_j\right)^2\\
				&+\frac{D_{xy}}{2}\sum_i\left[\left(S_i^x+iS_i^y\right)^6+\left(S_i^x-iS_i^y\right)^6\right]\\&-g\mu_{\text{B}}\sum_i\mathbf {S}_i\cdot \mathbf {H}\ .\label{Eq:SIHam}
			\end{aligned}
		\end{split}
	\end{equation}
	The model has nearest ($J_1$), next-nearest ($J_2$),  third-neighbor ($J_3$) effective exchange interactions, and a nearest-neighbor biquadratic interaction ($K_1$) along the $\mathbf {c}$ axis \cite{Pimpinelli_1989} that reproduces the $H=0$ broken-helix order. Each Eu layer is represented by a fixed-length spin $\mathbf {S}$ at position $i$ or $j$ along $\mathbf{c}$ and the spins are oriented in the $\mathbf {ab}$ plane by an implied large in-plane magnetic anisotropy. We also introduce a nearest-neighbor biquadratic exchange coupling $K_{1}>0$. $g$ is the spectroscopic splitting factor, $\mathcal{H}_{\text{in-plane}}$ accounts for the effective ferromagnetic interaction ($J_0$) within each Eu layer and the large in-plane anisotropy, and $D_{xy}$ is a six-fold magnetic anisotropy constant pinning the blue spins in Fig.~$1$ of the main text along the $[\bar{1}\,1\,0]$, $[\bar{1}\,\bar{2}\,0]$, or $[2\,1\,0]$ axis, depending on the domain being considered.
	
	The Luttinger-Tisza method gives the exact solution for our $J_1-J_2-J_3$ model given by Eq.~\eqref{Eq:SIHam} (with all other parameters set to zero); the ground state is a single-$q$ spiral with the periodicity determined by the ratio of the interaction parameters $J_1$, $J_2$, and $J_3$. For $J_1/J_3 \gtrsim J_2/J_3$, the ground state is an A-type antiferromagnet (AFM), with the ferromagnetic (FM) Eu layers stacked antiferromagnetically along the $c$-axis. In the opposite limit, for $J_1/J_3 \lesssim J_2/J_3$, the ground state is a regular $c$-axis helix. On the boundary between A-type and helical-magnetic order for $J_1/J_3 \approx J_2/J_3$ where the helical-turn angle is $\theta \approx 60$\degree, the ground state belongs to a highly degenerate manifold, containing additional lower-symmetry multi-$q$ interpolating states between the A-type and the regular-helix magnetic orders, including the broken-helix magnetic order.
	
	To stabilize the broken-helix magnetic order from the manifold of the degenerate states, we introduce the AFM biquadratic coupling $K_1>0$ between neighboring layers.  The AFM biquadratic coupling favors a perpendicular orientation of neighboring spins. Alternatively, the broken-helix state can be stabilized by other interactions that favor perpendicular orientations of the neighboring spins, such as the Dzyaloshinskii–Moriya (DM) interaction $\sim \mathbf{S}_i \times \mathbf{S}_j = (S_i^x S_j^y - S_i^y S_j^x) \hat{z}$. From our analysis, the DM interaction $\mathbf{D}^{\text{DM}}_{ij} \sim \mathbf{r}_i \times \mathbf{r}_j$ between the neighboring Eu atoms at positions $i$ and $j$ is allowed to exist in the compound, as the neighboring Eu atoms along the $c$-axis do not have an inversion center~\cite{Moriya_1960}. The difference with the biquadratic interaction $K_1$ is that the choice of the DM interaction fixes the chirality of the broken-helix magnetic order. Thus far, there is no experimental evidence to indicate whether one vector chirality is favored over the other.
	
	\begin{figure*}
		\centering    \includegraphics[width=\linewidth]{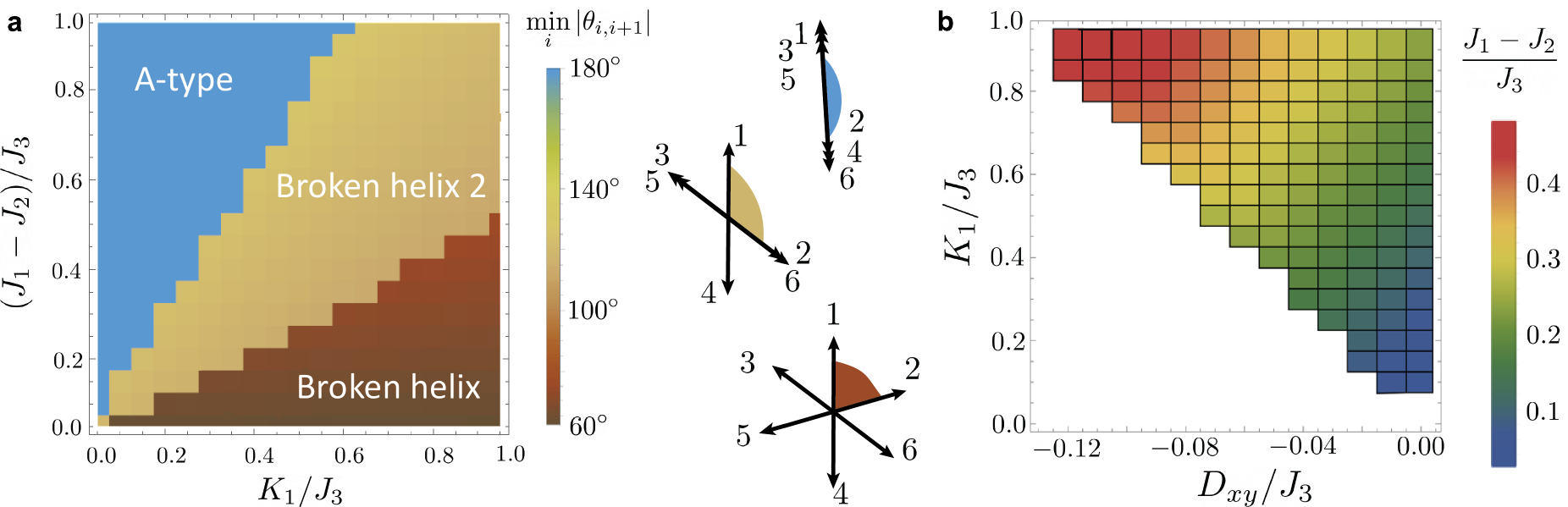}
		\caption{\textbf{Further results from analytic calculations modeling the magnetic order.} \textbf{a} The variational phase diagram for the model given in Supplementary Eq.~\eqref{MTI_Eq:J1-J2-J3-Kxy-Ham} for $D_z>0$, $D_z \gg |J_1|, |J_2|, |J_3|, |K_1|$, $J_3>0$, $K_1>0$.  The phases are parameterized by the minimal turn-angle between neighboring spins (Eu layers) $\min\limits_{i} |\theta_{i,i+1}|$ for a given spin configuration, which is shown by the color. The representative states to the right of the phase diagram show A-type magnetic order (blue) with $\min\limits_{i} |\theta_{i,i+1}| =180$\degree, the broken-helix magnetic order (dark brown) with $\min\limits_{i} |\theta_{i,i+1}| \in (\approx 62\text{\degree}, \approx 75\text{\degree})$ and the broken-helix-2 phase (beige) with $\min\limits_{i} |\theta_{i,i+1}| \in (\approx 112\text{\degree}, \approx 128\text{\degree})$. \textbf{b}~The variational phase diagram that illustrates the broken-helix magnetic order for the model given in Eq.~($1$) of the main text and  zero magnetic-field using the turn angles previously found by experiment \cite{Riberolles_2021}. The constraint on the angles fixes the $(J_1-J_2)/J_3$ ratio, with the value given by the color. The white region indicates the sets of parameters for which there are no solutions.
		}
		\label{App:PD}
	\end{figure*}
	
	We now show how the broken-helix magnetic order arises from the $J_1$-$J_2$-$J_3$-$K_1$-$D_z$ Heisenberg model with $K_1>0$ and $D_z$ a magnetic anisotropy constant. The model is given by the Hamiltonian
	\begin{equation} 
		\begin{split}
			\begin{aligned}
				\mathcal{H}_0' = \sum_{n=1}^3 J_n \sum_{\langle i, j \rangle_n} \mathbf{S}_i \cdot \mathbf{S}_j &+ K_1 \sum_{\langle i, j \rangle} (\mathbf{S}_i \cdot \mathbf{S}_j)^2\\ &+ D_z\sum_i (S_i^z)^2 \,. 
			\end{aligned}
		\end{split}\label{MTI_Eq:J1-J2-J3-Kxy-Ham}
	\end{equation}
	We solve this model by considering the variational minimization ansatz assuming the six-fold periodicity of the spin structure. The obtained phase diagram in the limits $D_z>0$, $D_z \gg |J_n|, |K_1|$, $n\in\{1,2,3\}$ that constrain the spins to lie in the $ab$ plane, and for $J_3>0$, $K_1>0$, is shown in Supplementary Fig.~\ref{App:PD}a, as a function of $K_1/J_3$ and $(J_1-J_2)/J_3$. The phases are specified by the minimal value of the relative turn angle between spins in nearest-neighbor layers, $\min \limits_i |\theta_{i,i+1}|$. Illustrated are the representative states from each of the phases. The phase diagram shows the stabilization of the six-fold periodic broken-helix state for $K_1/J_3>0$ and small $(J_1-J_2)/J_3 >0$. In what follows, we focus on the set of parameters that stabilize the broken-helix magnetic order.
	
	To model the experimentally observed pinning of a $C_2^\prime$ (i.e.\ $2^\prime$) high-symmetry axis of the broken-helix order to one of the three $C_2$ axes we additionally introduce a small six-fold easy-axis anisotropy $D_{xy}<0$, as allowed by the hexagonal symmetry of the lattice. The final considered model here is given in  Eq.~\eqref{Eq:SIHam}, but for zero magnetic field, and we have assumed a large implied in-plane anisotropy. Supplementary Figure~\ref{App:PD}b features a region of the stabilized broken-helix ground state as a function of the six-fold easy-axis anisotropy $D_{xy}/J_3<0$ and the AFM nearest-neighbor biquadratic coupling $K_1/J_3>0$. The solution is obtained variationally, assuming spin configurations with a periodicity of up to $12$ spins per unit cell.
	
	The set $\{D_{xy}/J_3<0,K_1/J_3>0\}$ fixes the value of $(J_1-J_2)/J_3 >0 $ for the realization of broken-helix order with the helix angles observed by experiment. This leaves an unconstrained parameter $(J_1+J_2)/J_3$, allowing, interestingly, the three combinations for the signs of $J_1/J_3$ and $J_2/J_3$, as long as  $(J_1-J_2)/J_3 >0 $ for the occurrence of the broken-helix with the specified turn angles and fixed $(J_1-J_2)/J_3 >0 $: ($1$) the AFM signs $J_1/J_3>0$, $J_2/J_3 > 0 $, ($2$) the FM signs $J_1/J_3<0$, $J_2/J_3 < 0$, and ($3$) mixed signs $J_1/J_3>0$ and $J_2/J_3<0$. While $(J_1+J_2)/J_3$ is a free parameter for the stabilization of the broken helix, it will play a role in the evolution of the broken-helix magnetic order in the magnetic field. The full analysis of the evolution of the broken helix ground state in a magnetic field across different interaction parameter regimes, including the discussion of the choice of parameters reported here, are given in Ref.~\cite{nedic2023effects}.
	
	\begin{figure*}
		\includegraphics[]{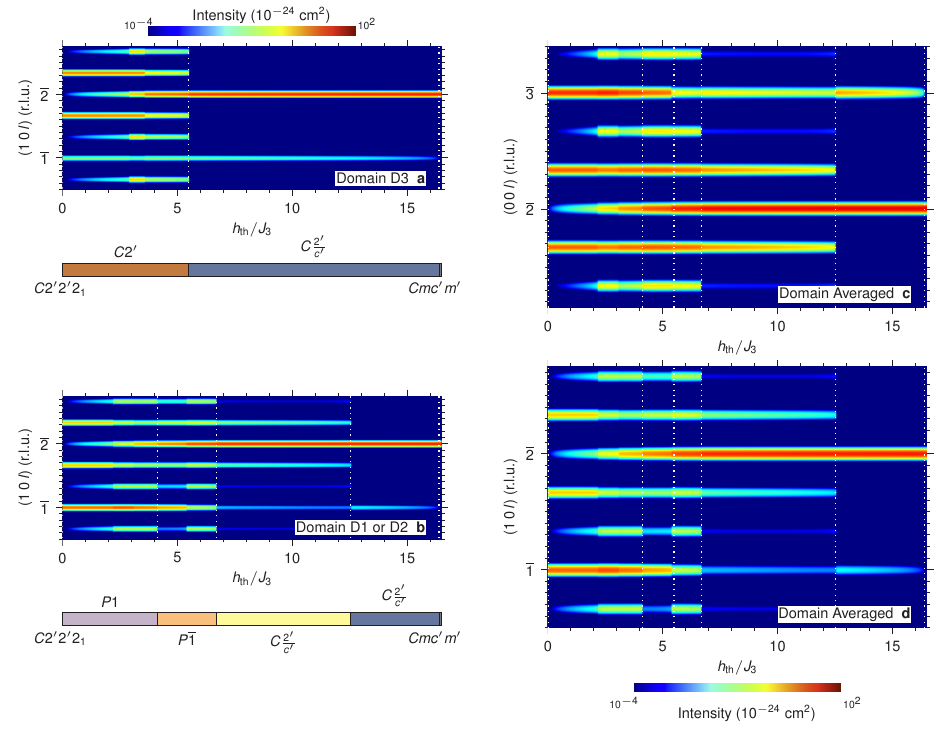}
		\caption{
			\textbf{Calculated magnetic neutron-diffraction patterns for $\bm{\phi_\text{rb}=122\degree}$.} \textbf{a},\textbf{b} Calculated $(1\,0\,l)$ magnetic diffraction patterns based on the XY Hamiltonian described in the text for a magnetic field $\mathbf{H}$ applied along the $b$ axis and a temperature $T=0$~K as a function of $h_{\text{th}}/J_3=g\mu_{\text{B}}H/J_3$ for domains $\text{D}3^\pm$ (a) and $\text{D}1^\pm$ or$\text{D}2^\pm$ (b). Dotted lines indicate magnetic phase transitions and bar graphs show the field regions for the different magnetic phases. Example magnetic structures for each field region are also shown, where numbers correspond to the Eu layers in Fig.~$1$a of the main text. The diagrams in a are for $\text{D}3^+$  and in b are for domain $\text{D}1^+$.  \textbf{c}, \textbf{d} Calculated $(0\,0\,l)$ (c) and $(1\,0\,l)$ (d) domain-averaged magnetic neutron-diffraction pattern as a function of $h_{\text{th}}/J_3$ for $T=0$~K. Intensities correspond to $1~\mu_{\text{B}}$ per Eu for a magnetic unit cell tripled along the c-axis.   
		}
		\label{Fig:SI_Theo_Diff_Field}
	\end{figure*} 
	
	The calculation results in Fig.~$4$ of the main text are done using the microscopic magnetic parameters given above which give $\phi_\text{rb}=127\degree$, the value we previously reported obtaining from experiment \cite{Riberolles_2021}. However, we also considered relaxing this condition and choosing $\phi_\text{rb}=122\degree$ instead, close to the value we found for the sample used in the present experiments.  The parameters for $\phi_\text{rb}=122\degree$ are  $J_1=0.9$, $J_2=0.8$, $K_{1}=0.2$, and $D_{xy}=-0.2$ for $J_3=1$. Supplementary Figure~\ref{Fig:SI_Theo_Diff_Field}a shows the calculated $(1\,0\,l)$ magnetic neutron diffraction pattern for $\text{D}3^\pm$ and Supplementary Fig.~\ref{Fig:SI_Theo_Diff_Field}b shows the pattern for $\text{D}1^\pm$ or $\text{D}2^\pm$. The domain-averaged patterns are shown in Supplementary Figs.~\ref{Fig:SI_Theo_Diff_Field}c and \ref{Fig:SI_Theo_Diff_Field}d for $(0\,0\,l)$ and $(1\,0\,l)$, respectively. Dotted white lines indicate phase transitions, and the extent of each magnetic phase and magnetic space group is shown by the bar graphs. The phases are the same as those shown in Fig.~$4$ of the main text but with different critical fields. Notably, the parameters used extend the field range over which the $\bm{\tau_1}$-Bragg peaks exist.

	\clearpage
	\section*{Supplementary Note 3}
	\subsection*{Combining broken helix domains}
	\begin{figure}
		\centering    \includegraphics[width=\linewidth]{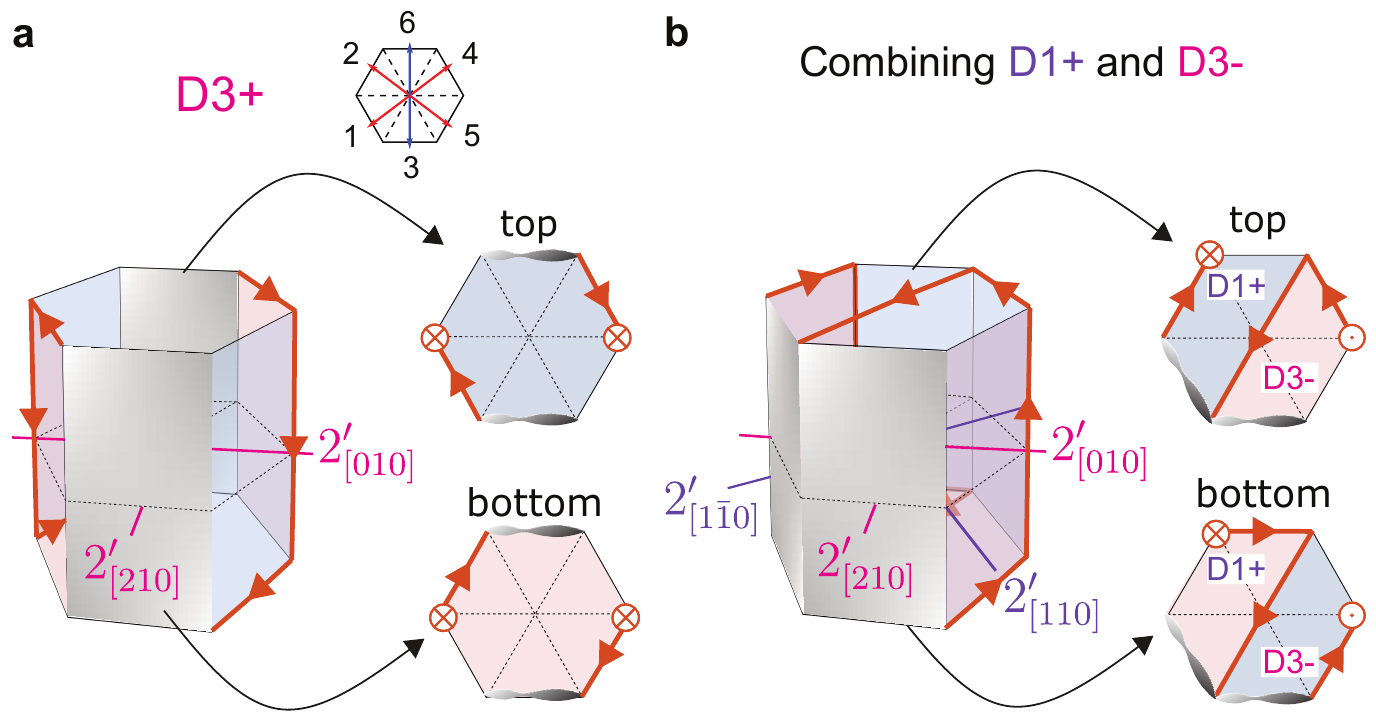}
		\caption{\textbf{Diagrams showing further examples of surface-Dirac and hinge states for different magnetic domains.} \textbf{a} Magnetic domain D$3^+$ with surfaces with a $2^\prime$ axis normal shaded light gray. Red and blue shadings indicate opposite signs for the effective mass associated with gapped surface-Dirac states. Orange paths indicate hinge states. \textbf{b} The intersection of domains D$1^+$ and D$3^-$ along the $[1\,2\,0]$ domain wall.  Symmetry-enforced hinge-CES pinned to the magnetic domain wall are present.
		}
		\label{FigSI:D3}
	\end{figure}
	
	\begin{figure*}
		\centering    \includegraphics[width=\linewidth]{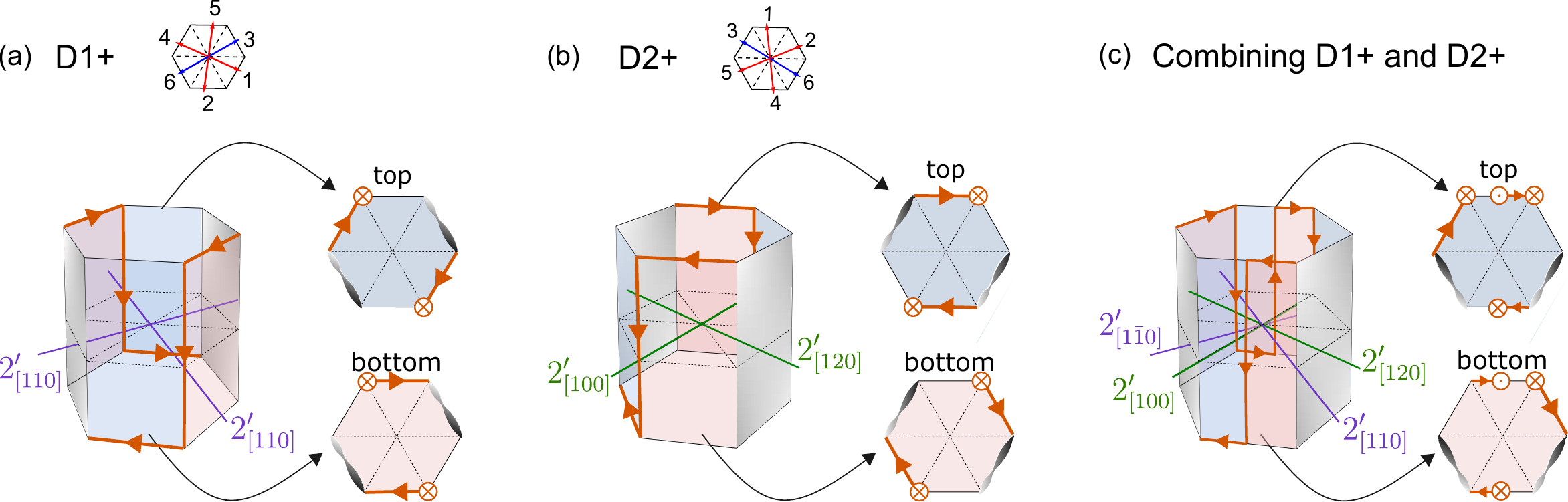}
		\caption{\textbf{Diagrams showing further examples of surface-Dirac and hinge states for different magnetic domains.} Magnetic domains \textbf{a} D$1^+$ and \textbf{b} D$2^{+}$ of the broken helix state. Surfaces normal to a $2^\prime$ axis normal are shaded light gray. Red and blue shadings indicate opposite signs for the effective mass associated with gapped surface-Dirac states.  \textbf{c} The intersection of domains D$1^+$ and D$2^+$ along a side surface.  Symmetry-enforced hinge-CES pinned to the magnetic domain wall are represented by the orange paths.
		}
		\label{FigSI:D12}
	\end{figure*}

	Supplementary Fig.~\ref{FigSI:D3}a shows a diagram of magnetic domain D$3^+$ with its $2^\prime$ symmetry axes. The topologically protected hinge states are shown by the orange paths, and red and blue shading indicates opposite signs for the effective mass term associated with gapped Dirac surface states. Supplementary Figure~\ref{FigSI:D3}b shows the intersection of D$1^+$ and D$3^-$ along the $[1\,2\,0]$ domain wall. Symmetry-enforced hinge states pinned to the magnetic domain wall are present. Supplementary Fig.~\ref{FigSI:D12} shows a case for the intersection of D$1^+$ and D$2^+$.

\end{document}